\documentclass[twocolumn]{aastex631}

\newcommand\blfootnote[1]{%
  \begingroup
  \renewcommand\thefootnote{}\footnote{#1}%
  \addtocounter{footnote}{-1}%
  \endgroup
}

\begin{document}

\title{A Detailed Look at a Trio of Changing-Look Quasars: Spectral Energy Distributions and the Dust Extinction Test$^{*\dagger}$}\blfootnote{$^*$ Based on observations obtained with the Hobby-Eberly Telescope (HET), which is a joint project of the University of Texas at Austin, the Pennsylvania State University, Ludwig-Maximillians-Universitaet Muenchen, and Georg-August Universitaet Goettingen. The HET is named in honor of its principal benefactors, William P. Hobby and Robert E. Eberly.}\blfootnote{$^\dagger$ Based on observations made with the NASA/ESA Hubble Space Telescope, obtained at the Space Telescope Science Institute, which is operated by the Association of Universities for Research in Astronomy, Inc., under NASA contract NAS5-26555. These observations are associated with program GO-14799.}

\author[0000-0003-1752-679X]{Laura Duffy}
\affiliation{Department of Astronomy and Astrophysics and Institute for Gravitation and the Cosmos, Penn State University, 525 Davey Lab, 251 Pollock Road, University Park, PA 16802, USA}

\author[0000-0002-3719-940X]{Michael Eracleous}
\affil{Department of Astronomy and Astrophysics and Institute for Gravitation and the Cosmos, Penn State University, 525 Davey Lab, 251 Pollock Road, University Park, PA 16802, USA}

\author[0000-0001-8557-2822]{Jessie C. Runnoe}
\affil{Department of Physics and Astronomy, Vanderbilt University, Nashville, TN 37235, USA}

\author[0000-0001-8665-5523]{John J. Ruan}
\affil{Department of Physics and Astronomy, Bishop’s University, 2600 College St., Sherbrooke, QC J1M 1Z7, Canada}

\author[0000-0002-6404-9562]{Scott F. Anderson}
\affil{Department of Astronomy, University of Washington, Box 351580, Seattle, WA 98195, USA}

\author[0000-0002-0489-1686]{Sabrina Dimassimo}
\affil{Department of Astronomy, Yale University, New Haven, CT 06520, USA}

\author[0000-0002-8179-9445]{Paul Green}
\affil{Center for Astrophysics, Harvard \& Smithsonian, 60 Garden Street, Cambridge, MA 02138, USA}

\author[0000-0002-5907-3330]{Stephanie LaMassa}
\affil{Space Telescope Science Institute, 3700 San Martin Drive, Baltimore, MD 21218, USA}




\begin{abstract}

Changing-look quasars exhibit dramatic variability in broad emission-line fluxes on short timescales. This behavior is challenging to many models of the quasar broad line region, due in large part to the short transition times between high and low states.  In order to constrain the cause of the dramatic variability, we obtained contemporaneous Hubble Space Telescope UV and Hobby Eberly Telescope optical spectra of three changing-look quasars caught in their low state. We use these spectra, along with archival spectra taken during both the high and low states, to investigate potential scenarios for the change in state. Our data strongly disfavor a variable dust obscuration scenario for these three CLQs, and instead suggest that the observed transformation reflects a change in the intrinsic luminosity of the central engine. We also find that the low-state spectral energy distributions of all three quasars are reminiscent of those of low-luminosity active galactic nuclei, which suggests that the transition may result from a change in accretion flow structure caused by a reduced Eddington ratio.

\end{abstract}

\keywords{}


\section{Introduction} \label{sec:intro}
Changing-look quasars (CLQs), first identified by \cite{lamassa15}, are luminous active galactic nuclei (AGN) that show dramatic changes in both their broad emission lines and their non-stellar continuum levels on month-to-year timescales \citep[see][for a review]{ricci23}. These changes are marked by a transition between quasar-dominated and host galaxy-dominated spectra. In the `high state', the quasars exhibit both broad and narrow emission lines and a strong non-stellar continuum, while in the `low state', the broad lines and non-stellar continuum are considerably weaker. This sort of transition was first detected in the optical spectra of lower luminosity Seyfert galaxies \citep[e.g.][]{goodrich89, goodrich90, goodrich95}. These galaxies either transitioned from having weak or undetectable broad H$\beta$ to having strong broad H$\beta$ emission \citep{goodrich90}, or the reverse \citep{goodrich95}. Quasars have also long been known to exhibit continuum variability on both short and long timescales \citep[e.g.][]{giveon99, vandenberk04, wilhite05, macleod12}. In the UV and optical, this means that the quasar spectral energy distribution (SED) is generally `bluer-when-brighter' \citep[e.g.][]{giveon99, vandenberk04, wilhite05, ruan14, macleod16}, and broad emission lines lag behind the continuum in their responses to flux changes due to the light travel time from the ionizing source \citep[e.g.][]{peterson93, grier12, shen15, homayouni19, sharp24}. Historically, the phrase `changing-look' has also been used to describe changes observed in the X-ray spectra of AGN \citep[e.g.][]{matt03, piconcelli03, marchese12, ricci16}, but here we refer only to behavior in the ultraviolet (UV), optical, and infrared (IR).


Turn-off CLQs (those that transition from a high state to a low state) are frequently identified from follow-up photometric and spectroscopic observations of objects previously identified as quasars \citep[e.g.][]{lamassa15, macleod16,runnoe16, ruan16, yang18, green22, zeltyn22, zeltyn24}. Particularly in the turn-on case, CLQs have also been identified by dramatic changes in their optical light curves and confirmed through repeat spectroscopy \citep[e.g.][]{macleod16, gezari17, yang18, ln22, wang24, yang24}. Both recent large time-domain surveys and more targeted searches have led to the discovery of hundreds of CLQs, but the physical mechanisms behind the transition remain elusive \citep[e.g.][]{macleod19, graham20, guo24, guo24_2, wang24, zeltyn24}. Studies have found that CLQs may have preferentially lower Eddington ratios than the general population of quasars \citep[e.g.][]{macleod19, green22, zeltyn24}.


CLQs represent a valuable means through which to study the properties of the supermassive black hole (SMBH) itself, the structure of the accretion flow, and the properties of quasar host galaxies \citep{charlton19, dodd21}. Such dramatic changes in quasars challenge our current models of their structure and accretion flows. A number of different scenarios have been proposed to explain the observed transitions, but the relatively short transition times \citep[see][for example]{trakhtenbrot19} and the sometimes recurrent nature of the phenomenon \citep[see][for example]{zeltyn22} challenge models. Many studies that investigate variability in AGN focus on two main mechanisms; variable accretion rates between states \citep[e.g.][]{lamassa15, runnoe16, ruan16, yang18, ruan19, green22}, and variable dust obscuration of the ionizing source \citep[see][]{maiolino10, markowitz14, zeltyn22}. Other potential mechanisms have been proposed \citep[e.g. tidal disruption of a star,][]{merloni15}, but these explanations are less likely in most cases \citep[see][]{ruan16, yang19}. 

Dramatic variations in the accretion rate are often cited in works that compare the state transitions seen in CLQs to those seen in galactic X-Ray binaries \citep[XRBs; see][]{sobolewska11, ruan19, jin21, yang23}. These works suggest that the transformation may be associated with a change in Eddington ratio, $\eta_\mathrm{Edd}$, and thus a change in the accretion flow structure. In these models, when $\eta_\mathrm{Edd}$ drops below a certain value (generally $\sim$ 0.01), the structure of the accretion flow changes from a standard thin disk \citep{shakura73} to an radiatively inefficient or advection dominated accretion flow \citep[RIAF or ADAF; e.g.][]{narayan95} with or without a luminous jet \citep{markoff01}. 

The short transition times between high and low states remains challenging in this framework, however. In the standard model for thin accretion disks \citep{shakura73}, the timescale for the luminosity to change as a result of large changes in the accretion rate is thought to be set by the viscous time, which is much longer than the observed transition time of CLQs. \cite{hameury09} applied models of the types of instabilities seen in XRBs (analogous to the ``dwarf nova'' instability that operates in the accretion disks of cataclysmic variables), which act on the thermal timescale, to AGN disks, and found variations on the scale of thousands to millions of years. Later works developed modifications to the standard thin disk model \citep[e.g.][]{dexter19} or adopted models based on galactic XRBs harboring stellar-mass black holes \citep[e.g.][]{noda18, sniegowska20} that significantly shorten the timescale for accretion disk variability. In particular, works that attribute accretion disk variability to a radiation pressure instability between an inner ADAF and an outer accretion disk or a magnetic-pressure dominated disk with outflows find timescales for changing-look behavior that are consistent with observations \citep[see][]{noda18, sniegowska20, feng21, wugu23}.


Variable dust extinction is the other commonly discussed explanation, motivated by the fact that X-ray observations have found dramatic changes in the obscuration of the central ionizing source \citep[e.g.][]{matt03, piconcelli03, marchese12, ricci16}. Moreover, \cite{goodrich90} found that changes in dust reddening provided a good explanation for the variability of Seyfert 1.8 and 1.9 galaxies. It is tempting to draw an analogy between these changing-obscuration X-ray AGN, variable Seyfert galaxies, and the behavior seen in changing-state quasars. Intervening dust clouds are not a completely satisfying explanation, however, both because the cloud of dust must be large enough to obscure the entire BLR, and because the crossing time for a cloud of dust orbiting the BLR is much longer than observed transition times \citep{lamassa15}. The problem can be avoided if variation in the UV continuum leads to the formation and sublimation of dust on very short timescales \citep{lamassa15, zeltyn22}. 

We thoroughly test the dust obscuration model for three previously identified CLQs. In this work, we present contemporaneous UV and optical spectra of each quasar in its low state, and compare with archival Sloan Digital Sky Survey (SDSS) optical spectra of each quasar in its high and low states. Using detailed spectral decomposition, we examine whether variable extinction models provide self-consistent explanations for the state changes. We also explore the low-state SED for each quasar. Previous tests of the variable dust extinction hypothesis have not made use of contemporaneous rest-frame UV and optical spectra.

We describe our target selection and observations in Section \ref{sec:obs}. In Section \ref{sec:meas}, we explain our measurement methods and in Section \ref{sec:indiv} we discuss measurements of individual quasars. In Section \ref{sec:analysis}, we conduct a multi-faceted test of the dust extinction hypothesis and we study the low-state SEDs. We discuss and interpret our results in Section \ref{sec:disc}. Finally, we summarize our work in Section \ref{sec:summary}. Throughout this work, we adopt a $\Lambda$CDM cosmology with $H_0 = 70$ km~s$^{-1}$ and $\Omega_m=0.30$.

\begin{deluxetable*}{ccccccc}
\tablecaption{Intrinsic CLQ Properties}
\tablewidth{0pt}
\label{table:props}
\setlength{\tabcolsep}{8pt}
\tablehead{
{SDSS} &{Truncated} & {} & {Luminosity} & {Angular Diameter} & {} \\
{Identifier} &{Identifier} & {\textit{z}}& {Distance (Mpc)} & {Distance (Mpc)} & {log($\frac{M_{BH}}{M_\odot}$)}\\
{(1)} & (2) & (3) & (4) & (5) & (6)}
\startdata
{J101152.98+544206.4} & {J1011} & {0.246} & {1240} & {800} & {7.6}\\
{J102152.34+464515.6} & {J1021} & {0.204} & {1000} & {690} & {8.3}\\
{J233602.98+001728.7} & {J2336} & {0.243} & {1220} & {790} & {8.1}\\
\enddata
\tablecomments{Column 1: Full SDSS object name, Column 2: Truncated name used in this paper, Column 3: Redshift, Column 4: Luminosity Distance (Mpc), Column 5: Angular Size Distance (Mpc), Column 6: Values for black hole mass are taken from \cite{runnoe16}, \cite{macleod16} [who utilize reported values from \cite{shen11}], and \cite{ruan16} respectively.} 
\end{deluxetable*}

\begin{deluxetable*}{ccccccc}
\tablecaption{Log of CLQ Observations}
\tablewidth{0pt}
\label{table:dates}
\setlength{\tabcolsep}{4pt}
\tablehead{
{}& {SDSS}& {SDSS} & {HET} & {HST} & {MDM}\\
{} &{high} & {low} & {low} & {low} & {low}\\
{Object} &{state} & {state} & {state} & {state} & {state}\\
{(1)} & (2) & (3) & (4) & (5) & (6)}
\startdata
{J1011} & {52652} & {57073} & {58231} &{58214} & {58199}\\
{} & {2003-01-13} & {2015-02-20} & {2018-04-23} &{2018-04-06} & {2018-03-22}\\
\\
{J1021} & {52614} & {56769} & {58226} &{58206} & {58199}\\
{} & {2002-12-06} & {2014-04-22} & {2018-04-18} &{2019-03-29} & {2018-03-22}\\
\\
{J2336} & {52096} & {55449} & {58044} &{58049} & {58010}\\
{} & {2001-07-06} & {2010-09-10} & {2017-10-18} &{2017-10-23} & {2017-09-14}\\
\enddata
\tablecomments{Column 1: Truncated SDSS object name, Columns 2--6: Modified Julian Date and calendar date (year-month-day) of observations. In the high state broad H$\beta$ is obvious, while in the low state broad H$\beta$ has diminished almost entirely.} 
\end{deluxetable*}

\section{Targets \& Observations}\label{sec:obs}

\subsection{Target Selection}

We selected three CLQs (SDSS objects J1011, J1021, and J2336 -- see Table \ref{table:props} for full identifiers) for follow-up study. All three were previously found as `turn-off' CLQs using repeat SDSS observations \citep[for discovery, see][]{runnoe16, macleod16, ruan16} and were included in the SDSS Data Release Seven quasar catalog (DR7Q), described in \cite{schneider10}. Because our tests rely on UV observations, we selected these three quasars to maximize the chance of observing any potential broad Ly$\alpha$ component in the UV spectra. These quasars were selected for their high [O~III] flux and the fact that their broad H$\alpha$ is still detectable in the low state. The CLQs span a range of properties -- their masses span a range $\log(M_{BH}/M_\odot)=7.6$--8.3, their high-state Eddington ratios span the range $\sim\,$0.005--0.15, and the factor by which their H$\alpha$ flux decays spans the range from $\sim\,$2--50 \citep[for details on mass, Eddington ratio calculations, see][]{runnoe16, macleod16, ruan16}.

Our data set includes optical spectra taken in the high and low states by the SDSS and contemporaneous low-state optical and UV spectra taken by the Hobby-Eberly Telescope (HET) second generation Low-Resolution Spectrograph (LRS2) and the Hubble Space Telescope (HST) Cosmic Origins Spectrograph (COS) respectively. We also use $r$-band images of J1011 and J1021, and V-band images of J2336 obtained by Jules Halpern at the MDM Observatory to supplement our data set. Table \ref{table:props} lists the full SDSS names of each quasar and the truncated identifier used in this paper, and includes information about the properties of each of the quasars. Table \ref{table:dates} lists the Modified Julian Date (MJD) of each of the observations referenced here. 

\subsection{Sloan Digital Sky Survey Optical Spectroscopy}\label{ssec:SDSS}
All SDSS spectra were taken with the SDSS 2.5 m telescope at Apache Point Observatory \citep{gunn06}. SDSS spectra were taken with two different instruments -- the SDSS spectrograph and the Baryon Oscillation Spectroscopic Survey \citep[BOSS,][]{dawson13, smee13, dawson16} spectrograph. Early high-state spectra cover a wavelength range 3800--9200~\AA\ and were taken with a $3\farcs0$ fiber with a resolving power $R\sim 2000$ \citep{york00, smee13}. The low-state spectra were generally taken with the upgraded BOSS spectrograph. BOSS spectra were taken with fibers of diameter $2\farcs0$ and cover a range 3600--10000~\AA\ at $R\sim2000$. 

J1011 and J1021 were first observed in SDSS-I/II and were later re-observed in a low state as part of the Time Domain Spectroscopic Survey \citep[TDSS,][]{morganson15,macleod18}, an effort in SDSS-IV intended to characterize variable objects through repeat spectroscopy.  

J2336 was observed four times with the SDSS spectrograph in SDSS-I/II. Following the discussion in \cite{ruan16}, we co-add the four early spectra to achieve a higher signal-to-noise ratio (SNR). The later spectrum was taken during SDSS-III, also with the SDSS spectrograph \citep{eisenstein11}.

\subsection{Hubble Space Telescope UV Spectroscopy}\label{ssec:HST}
We obtained low-state UV spectra of the three CLQs using the COS instrument on the HST \citep{coshand, cos} with the G140L and the G230L gratings. The HST data presented in this article were obtained from the Mikulski Archive for Space Telescopes (MAST) at the Space Telescope Science Institute. The specific observations analyzed can be accessed via \dataset[doi: 10.17909/nqv4-7f97]{https://doi.org/10.17909/nqv4-7f97}. The observations were part of program GO-14799 (P.I. Eracleous).
G140L covers the wavelength range 1150--1730~\AA\ at $R\sim 1500$--4000. The G230L grating covers the wavelength range 1570--3180~\AA\ at $R\sim2100-3900$. The COS entrance aperture has a redius of  $1\farcs25$. The G140L grating allowed us to observe both Ly$\alpha$ and C~IV emission in the UV. We used two central wavelength settings of the G230L grating, centered at 3000 and 3360~\AA\, in order to try to detect the C~IV, He~II and C~III] emission lines.

\subsection{Hobby-Eberly Telescope Optical Spectroscopy}\label{ssec:HET}
Contemporaneously with the HST spectroscopy, we obtained new low-state optical spectra of the three CLQs using the HET. \citep[][]{ramsey98, hill21}. These spectra were taken within 20 days of the HST observations with the red and blue channels of LRS2 \citep[LRS2-R and LRS2-B;][]{chonis16}. LRS2 uses a bundle of hexagonal fibers that cover a large area on the sky ($\sim 12''\times6''$). Each fiber has a diameter $\sim0\farcs6$. The bundle feeds either LRS2-B or LRS2-R, both of which are double spectrographs. The two arms of LRS2-B (named ``UV" and ``orange") cover a wavelength range 3640--7000~\AA\ at $R\sim1150$. The LRS2-R arms (names ``red" and ``farred") cover a wavelength range 6430--1056~\AA at $R\sim2500$. Here, we made use of the UV and orange channels of LRS2-B and the red channel of LRS2-R, and produced spectra that covered an observed frame wavelength range 3640--8450~\AA.

\subsection{MDM Observatory Imaging}\label{ssec:MDM}
We utilized new photometric observations of all three CLQs in the low state using the Templeton CCD on the Mcgraw-Hill 1.3 m telescope located at MDM Observatory. The Templeton CCD has a pixel scale of $0\farcs5$~pixel$^{-1}$, which corresponds to $\sim2.5$~kpc~pixel$^{-1}$ at the distance of our galaxies. We obtained observations of J1011 and J1021 in the SDSS $r$-band filter, and J2336 in the Johnson V-band filter. Under the conditions of the observations, the 1.3 m telescope delivered a point spread function (PSF) full-width at half maximum (FWHM) of $\sim4$ pixels, which corresponds to $\sim2\farcs0$.

\section{Data Processing and Measurements}\label{sec:meas}
\subsection{HST Data Processing}
Raw COS data were processed using the \texttt{calcos} pipeline, which corrects for instrumental effects, calibrates the wavelength scale, and extracts and produces final, one-dimensional, flux-calibrated spectra \citep{coshand}. We further processed COS data by combining multiple subexposures, weighted by their inverse squared errors such that the subexposures with the higher SNR were more heavily weighted. Finally, because the COS spectral resolution element is oversampled, we binned the spectra so that there were 20 pixels per bin. We then corrected for Galactic extinction, using the law defined by \cite{fitzpatrick99} and assuming R$_\mathrm{V} = 3.1$. 

We measured the SNR in the continuum of the binned G140L observations at 1350~\AA{}. All three observations have similar SNRs in the continuum. On average, they achieve an SNR of $\sim 2$.

\subsection{HET Data Processing}
The LRS2 data were initially processed with \texttt{Panacea}\footnote{\url{https://github.com/grzeimann/Panacea}}, which carries out bias subtraction, dark subtraction, fiber tracing, fiber wavelength evaluation, fiber extraction, fiber-to-fiber normalization, source detection, source
extraction, and flux calibration for each channel. The absolute flux calibration comes from default response curves derived from periodic standard star observations and measures of the mirror illumination as well as the exposure throughput from guider images.

Some observations had multiple subexposures -- in those cases, we combined the subexposures, weighted by their inverse squared errors such that the subexposures with the higher SNR were more heavily weighted. Spectra from the two arms of both the LRS2-B and the LRS2-R spectrographs also overlap in a narrow wavelength region. For each quasar, we re-normalized all spectra from every filter to the level of the orange filter spectrum using the overlapping regions, and combined them into a single spectrum. Following the methods described by \cite{wade88} and \cite{osterbrock90}, we made templates to correct for telluric absorption bands from O$_2$ and H$_2$O from the spectra of standard stars that were observed within a few days of the target observation. We then corrected for the telluric absorption bands and continuous atmospheric absorption. Finally, we corrected for Galactic extinction, again using a \cite{fitzpatrick99} extinction law and assuming R$_\mathrm{V} = 3.1$.

In order to make reliable flux comparisons between the high and low-states, we rescaled the low-state HET spectra to archival low-state SDSS spectra. We did this by measuring the [O~III] doublet flux in both the SDSS low-state and HET low-state observations and then re-scaling the HET observation such that the [O~III] flux matched the level from the low-state SDSS observation. This step accounted for differences in the absolute flux calibration between the HET and the SDSS spectrographs, and for the difference in the aperture sizes of the two instruments.

\subsection{Optical Photometry}
Using the MDM 1.3m images, we performed aperture photometry on the point sources at the centers of the CLQ host galaxies to get a constraint on the flux of each quasar and confirm that they remained in their low states. We first subtracted the background from each image. To determine the background level, we defined five apertures in the field, each with a radius of 10 pixels, averaged the count rate per pixel in each aperture, and then took the median of the average values from each aperture. We used stars in the same field to characterize the PSF of the MDM observations. To calibrate the flux scale of the images of J1011 and J1021, we identified three stars in the same field as the CLQ, placed apertures that were equivalent to twice the PSF FWHM of the observation around the star, and calculated the counts for each star in each aperture. We then repeated the process with sky-subtracted SDSS photometry, and calculated the SDSS flux of the same three stars. We used the median ratio between MDM flux and SDSS flux as the conversion factor for the MDM flux calibration.

To calibrate the V-band flux for J2336, we followed largely the same process with sky subtraction, PSF characterization, and reference star identification. We then transformed SDSS $g$- and $r$-band photometry of the three reference stars in the same field as the CLQ to the V-band using the \cite{jester05} transformation equations for stars. 

\begin{figure*}
\centering
\includegraphics[width=0.32\textwidth]{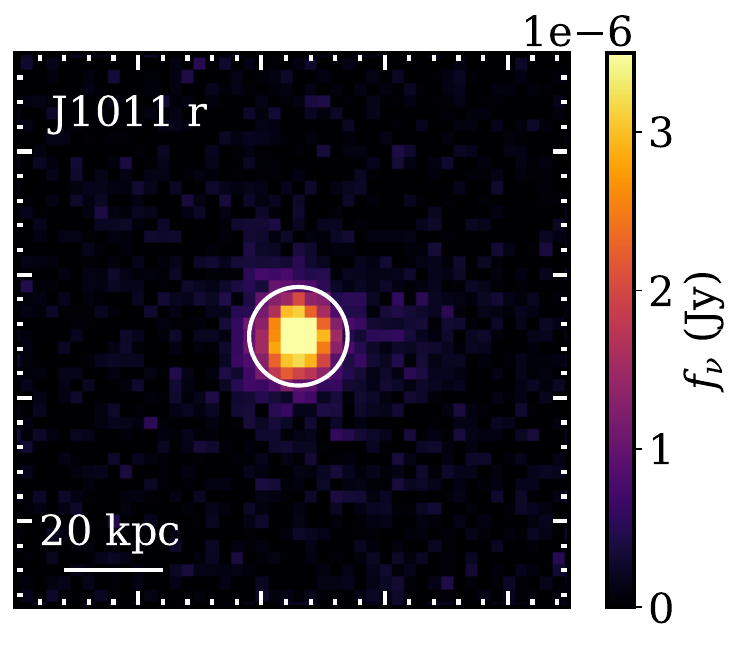}
\includegraphics[width=0.32\textwidth]{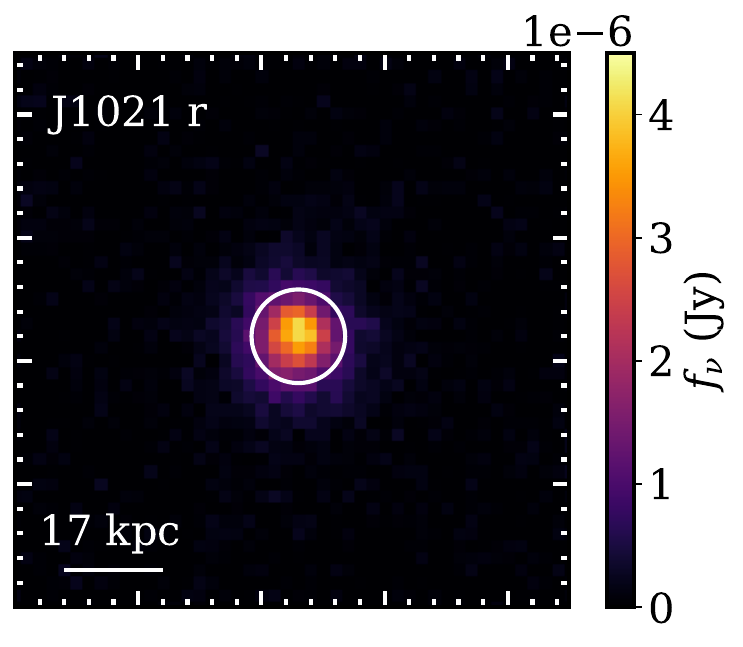}
\includegraphics[width=0.32\textwidth]{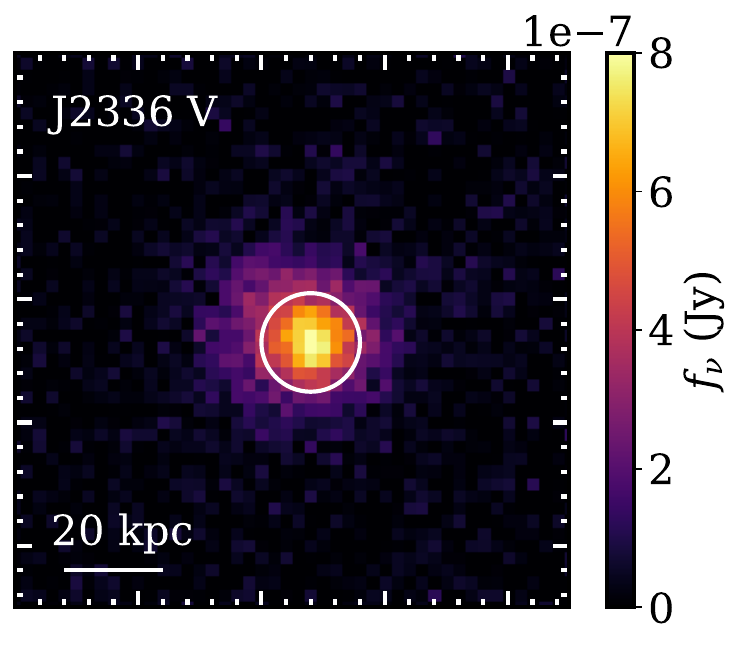}
\caption{MDM images of the three CLQs. Plotted on top in white is an aperture that represents the PSF 90\% light radius for each observation. We integrate the flux in that aperture and take that as the $r$ or V-band flux from the unresolved quasar within the larger extended galaxy. All three CLQs remained in their low states at the time of observation.}
\label{fig:phot}
\end{figure*}

In the calibrated MDM images, we placed an aperture with a diameter that encompasses 90\% of the light from an unresolved source on the center of the quasar, and measured the flux density in that region. Because we captured some host galaxy light within the aperture, the flux we measured from that aperture is strictly an upper limit to the flux from the quasar. While software tools exist that can spatially decompose the central point source from the extended galaxy in order to get more accurate photometric measurements, these upper-limits are sufficient here to confirm that the CLQs remained in their low states. We show the images of the quasars and their apertures in Figure \ref{fig:phot}.

Finally, we created synthetic photometry by multiplying the HET and SDSS spectra by the transmission curves of the filters used in the MDM imaging observations ($r$-band for J1011 and J1021 and V-band for J2336). We checked that synthetic magnitudes from the high-state spectra agreed well with the directly observed SDSS magnitudes. In the MDM observations, we found that the flux in the $r$-band for J1011 had dropped by a factor of $\sim2.2$ since the SDSS photometric observation was taken in 2002. The simulated $r$-band photometry in the high state was $\sim3$ times higher than the value we calculated for the low state. For J1021, the flux in the MDM $r$-band observations had also dropped by a factor of $\sim2.2$ compared to both the SDSS photometric observation from 2002 and the simulated value from the SDSS high-state spectrum. Finally, we found that J2336 had also dropped in flux since the SDSS photometry by a factor $\sim 1.5$. In all three cases, we confirmed that the quasars had remained in the low state since initial identification.

\subsection{Optical Spectral Decomposition}
We decomposed both the high and low-state spectra using the Python implementation of \texttt{pPXF} \citep{pPXF}. The standard version \texttt{pPXF} fits low-state spectra, where light from the host galaxy dominates, well but when there is significant contribution from the quasar, additional templates are required to achieve a good fit. To improve the ability of \texttt{pPXF} to fit the high-state spectra, we added templates to represent quasar power-law continua that follow a form $f_\lambda \propto \lambda^\alpha$ for a range of power-law indices, $-4<\alpha<2$ in steps of 0.1. We also included templates to model both UV and optical broad Fe~II emission \citep[following][]{veron04}. We created these templates using published Fe~II line wavelengths and strengths \citep{wills85}, and broadened them to a range of FWHMs from 500 to 11000~km~s$^{-1}$ in steps of 500~km~s$^{-1}$. To model other emission, we created templates representing contributions from higher-order Balmer lines \citep[following][]{storey95}, and the Balmer continuum \citep[following][]{balmercont1, balmercont2}. We only include higher-order Balmer line transition templates with upper levels $\mathrm{n}\leq50$ with a range of FWHM from 1000 to 11000~km~s$^{-1}$, again in steps of 500~km~s$^{-1}$. As a result, there is a gap in the resulting pseudo-continuum at $\lambda\sim3600$~\AA. Filling this gap was not necessary for our purposes \citep[it would have required including high-order Balmer transitions with upper levels up to $n\sim 400$; see discussion in][]{kovacevic14}. Further, we created Balmer continuum templates. We experimented with a range of different temperatures and found that the shape of the continuum was fairly insensitive to the temperature, and thus kept the temperature fixed at $\mathrm{T}=15000$ K. We varied the optical depth from $0.1<\tau<2.0$ in steps of $0.1$ and smoothed the final continuum templates with a Gaussian filter with $\mathrm{FWHM} = 4000$ km~s$^{-1}$. We used our implementation of \texttt{pPXF} to fit all of these components simultaneously, excluding the regions around emission lines.

First, we fit the low-state spectra. Once we removed starlight and any remaining quasar contributions from the spectra using \texttt{pPXF}, we were left with just the emission lines. Throughout this work, we used one or two Gaussian components to fit each emission line, treating narrow and broad emission lines separately, but we do not automatically attach any physical meaning to the individual Gaussian components. We limited the widths of narrow H$\alpha$ and H$\beta$ to be the same, and we tied the central wavelengths of those two narrow lines. In the low state, there were generally very weak to nonexistent broad H$\beta$ and symmetrical broad H$\alpha$ line profiles. Hence, we fit one broad Gaussian to the H$\alpha$ line profile, and then limited any potential broad H$\beta$ to have both the same velocity offset from the nominal position and FWHM as the broad H$\alpha$ component. With these constraints, we were able to determine the maximum contribution from broad H$\beta$ in this state. Once we had a fit, we bootstrap resampled each spectrum 1000 times. We did this by assuming the flux at each wavelength unit of the spectrum followed a normal distribution with an amplitude equivalent to the observed flux at that pixel and a standard deviation defined by the observed error at each pixel. We then resampled each pixel to create 1000 realizations of the spectrum, and performed each step of the fitting process again in order to obtain the uncertainties of the fit parameters.

Next, we fit the high-state spectra. In this state, a power-law continuum, Fe~II blended line complex, and Balmer continuum components were necessary to get a good fit. Although CLQ high-state fits frequently use the low-state stellar populations to constrain the high state, here we did not impose this constraint because the aperture of the SDSS fibers varies between observations, and also differs from that of the HET. Nonetheless, we compared the stellar populations fit by \texttt{pPXF} to the high state with the stellar populations from the low state. Although some level of difference in the stellar populations is to be expected due to the different aperture sizes, the differences were small. After removing the total continuum (including Fe~II) model, we fit the broad H$\alpha$ and H$\beta$ lines with two Gaussians. We initially allowed the narrow lines to be fit independently from high state values, but found a statistically insignificant difference between the high state and low state narrow emission strength, and thus we fixed the narrow line properties in the high state to those found in the low state. The narrow line region for quasars of this luminosity has typical sizes of $\sim 5$ kpc, or $\sim 2''$ \citep{bennert02}, and thus falls entirely within the aperture of both the SDSS fiber and the HET LRS2. Because of its size, the narrow line region also varies on much longer timescales than the broad line region, so we assumed that the flux in those lines remained mostly constant between observations. 

\begin{figure*}
\centering
\includegraphics[width=0.68\columnwidth]{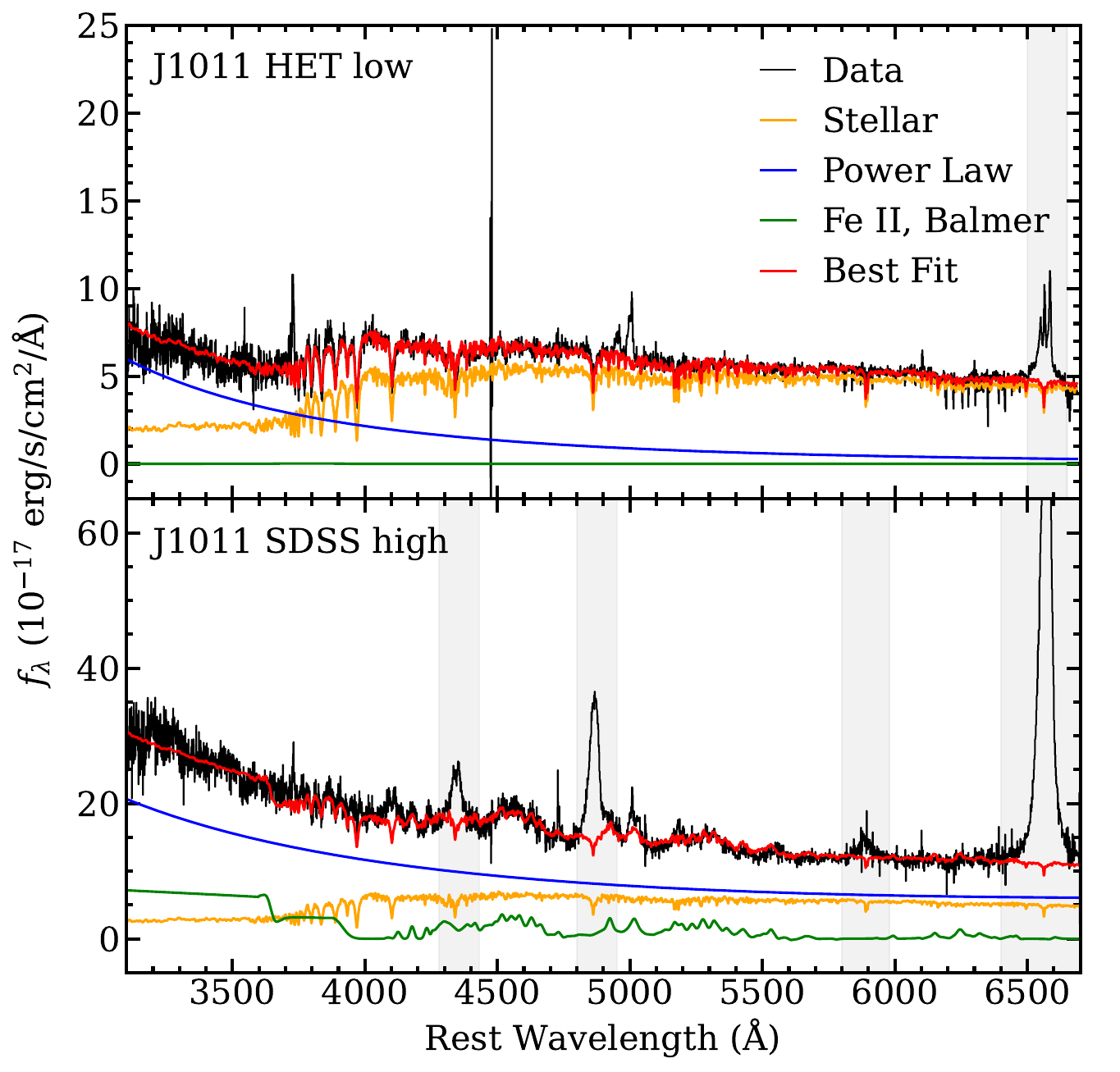}
\includegraphics[width=0.68\columnwidth]{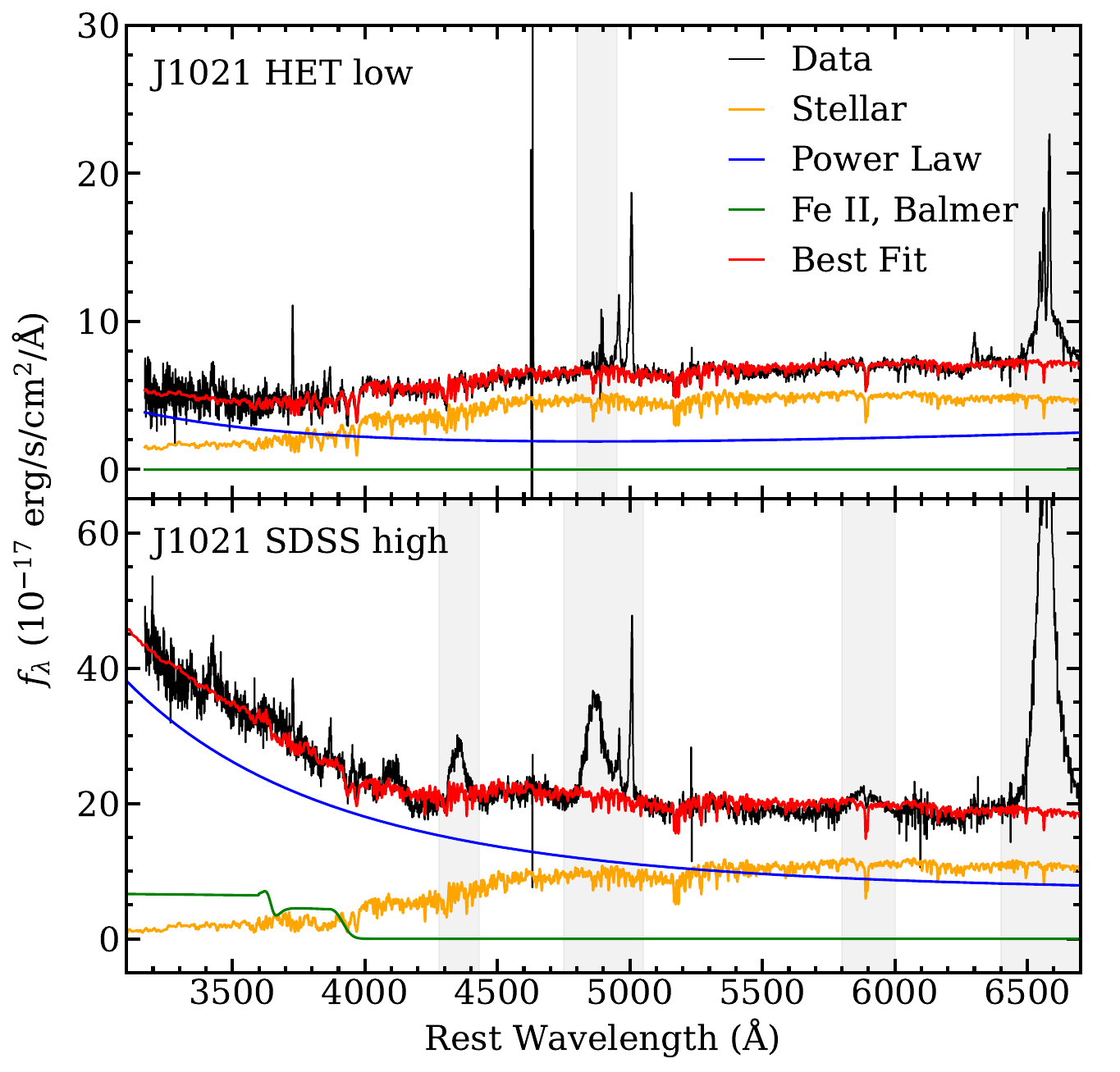}
\includegraphics[width=0.68\columnwidth]{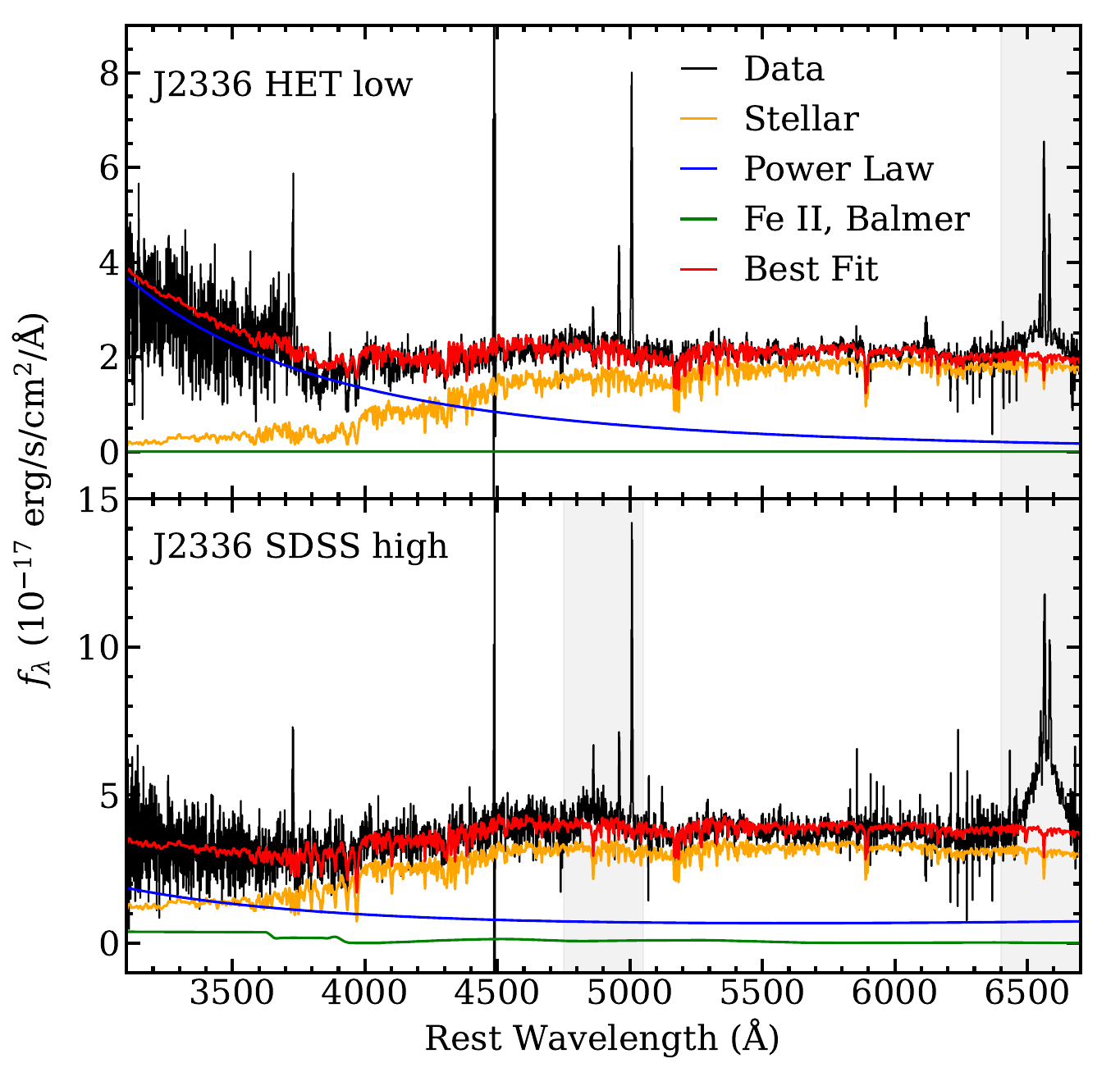}
\caption{Best-fit spectral decomposition for the low and high states of J1011, J1021, and J2336. We do not show broad and narrow emission line fits, but only the best-fit stellar contribution (orange), power law contribution (blue), Fe~II and Balmer continuum contributions (green), and the sum of those components (red). The regions blocked in light grey were not used for the fits, in order to avoid obvious broad emission lines. We removed the red, best-fit spectrum from the observed spectrum (black), and use the residuals to fit broad and narrow emission lines in the regions of interest.}
\label{fig:spec_fits}

\bigskip

\includegraphics[width=0.68\columnwidth]{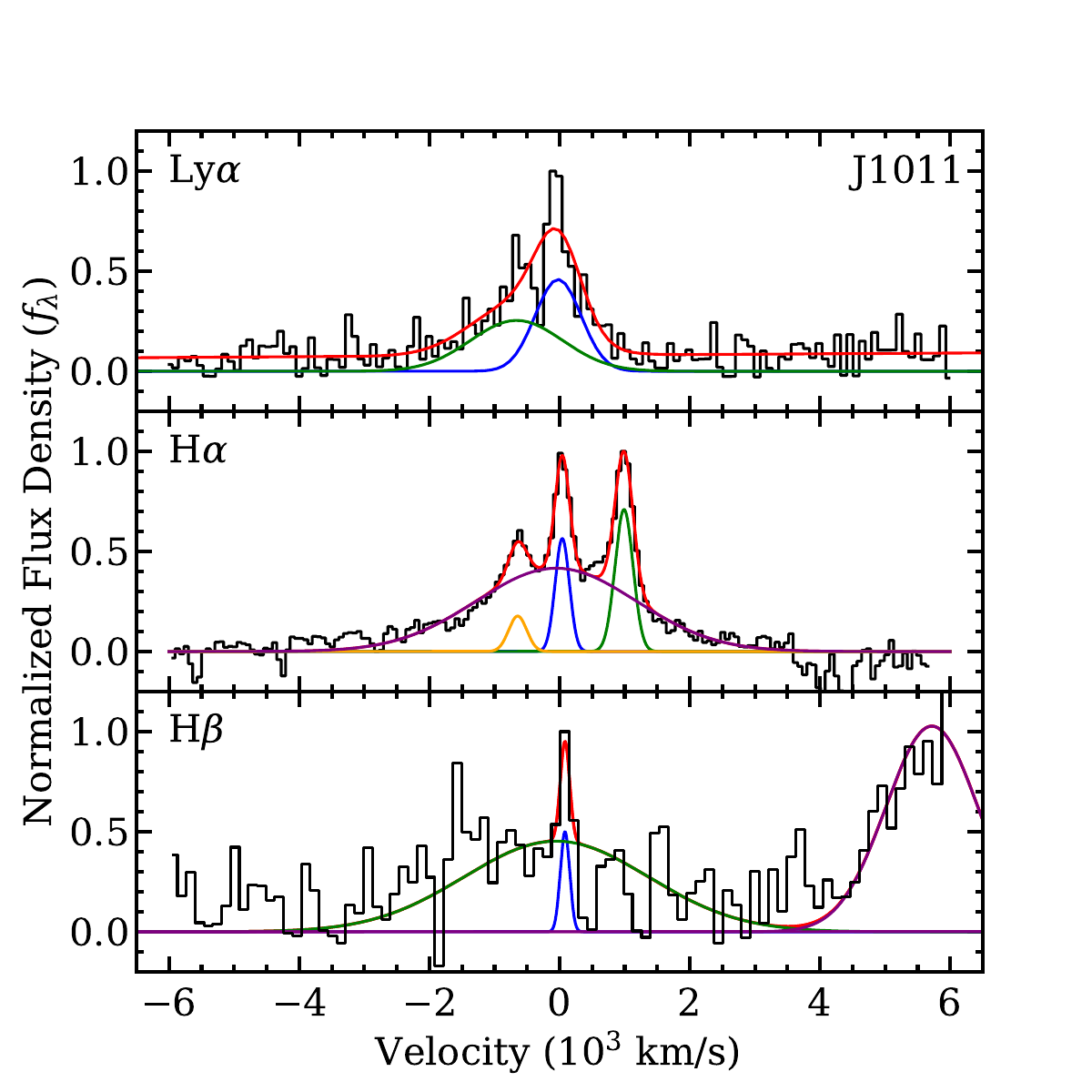}
\includegraphics[width=0.68\columnwidth]{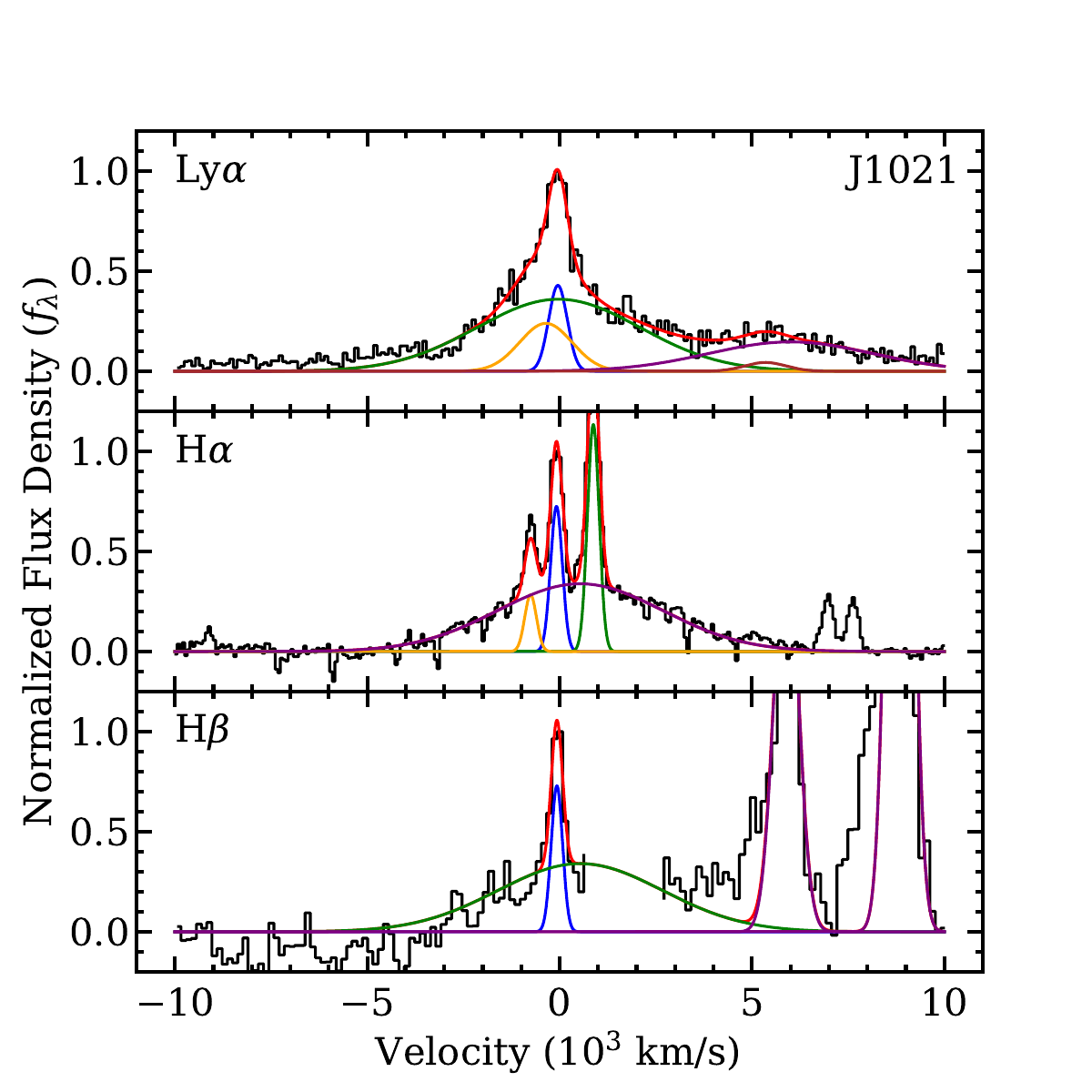}
\includegraphics[width=0.68\columnwidth]{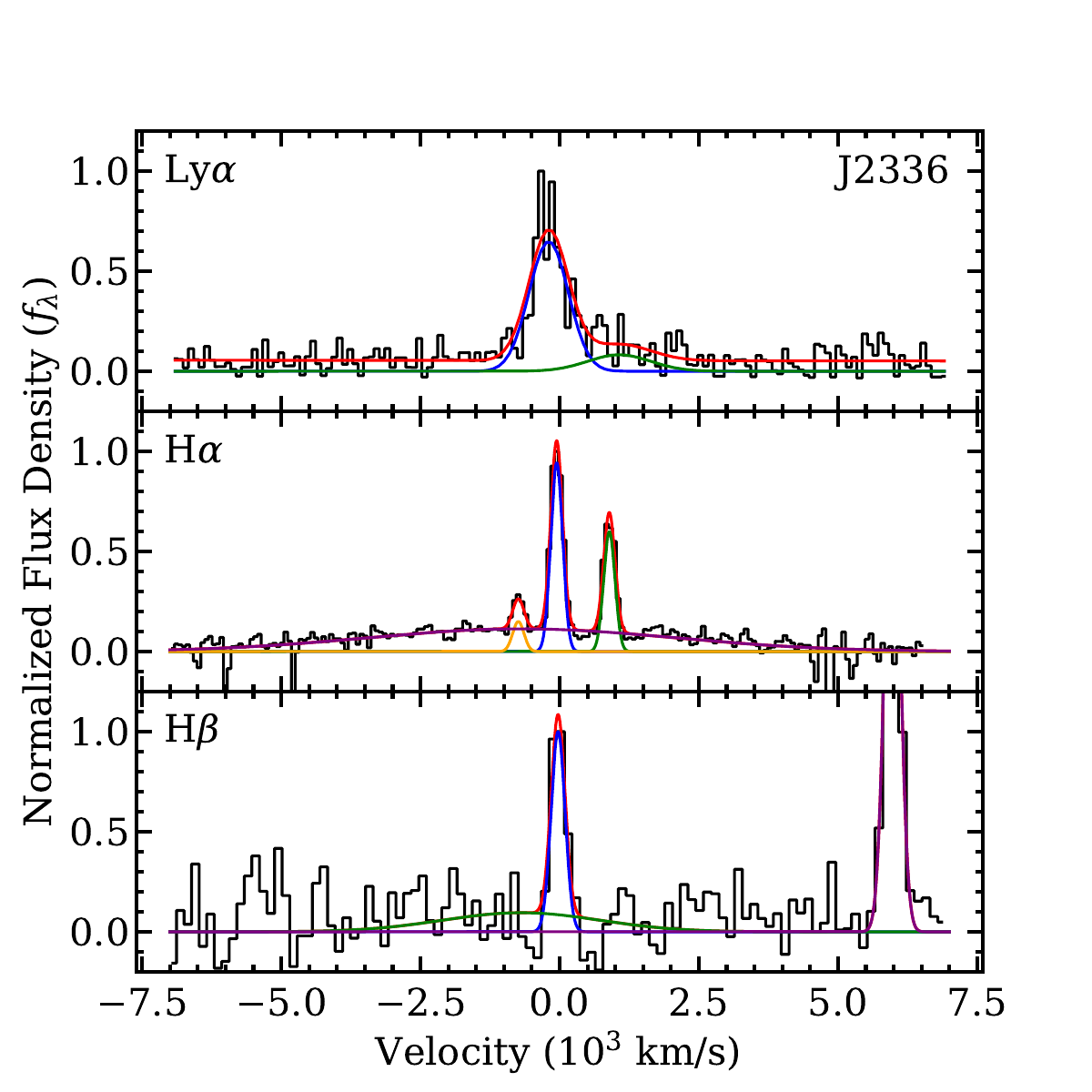}

\caption{Low-state Ly$\alpha$, H$\alpha$, and H$\beta$ emission line profiles for J1011, J1021, and J2336 respectively. We show the continuum-subtracted spectrum in black, the total best fit emission line profile in red, and various components of the fit in blue, green, and orange. In most cases, the emission lines are best fit with two components, a broad and a narrow one. In J1021, Ly$\alpha$ has both a more complex profile and contributions from N V, and so must be fit with three components for Ly$\alpha$, in green, blue, and orange, and two components for N V, in purple and brown. H$\alpha$ emission line profiles require more components due to [N II] emission. In the H$\beta$ profiles, we also fit each [O III] line with a single Gaussian, shown in purple. For J2336, the plotted green `broad' H$\beta$ represents the maximum contribution of H$\beta$ possible.}
\label{fig:emission_fits}

\bigskip

\includegraphics[width=0.68\columnwidth]{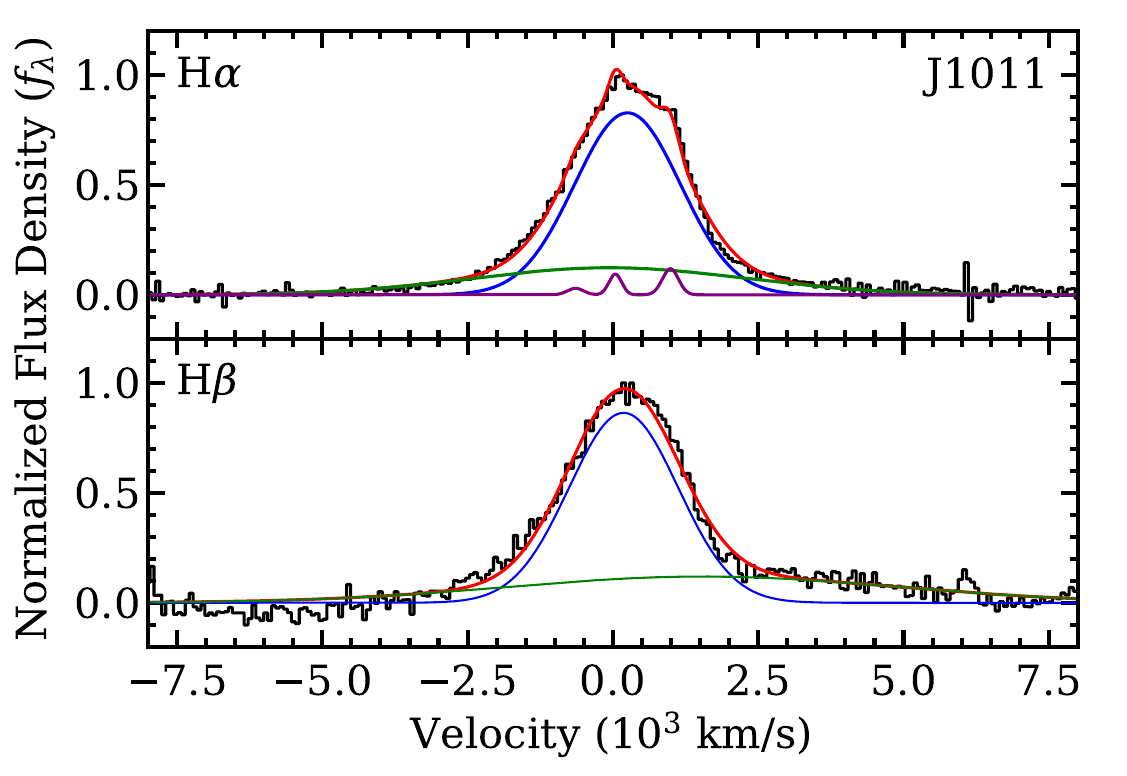}
\includegraphics[width=0.68\columnwidth]{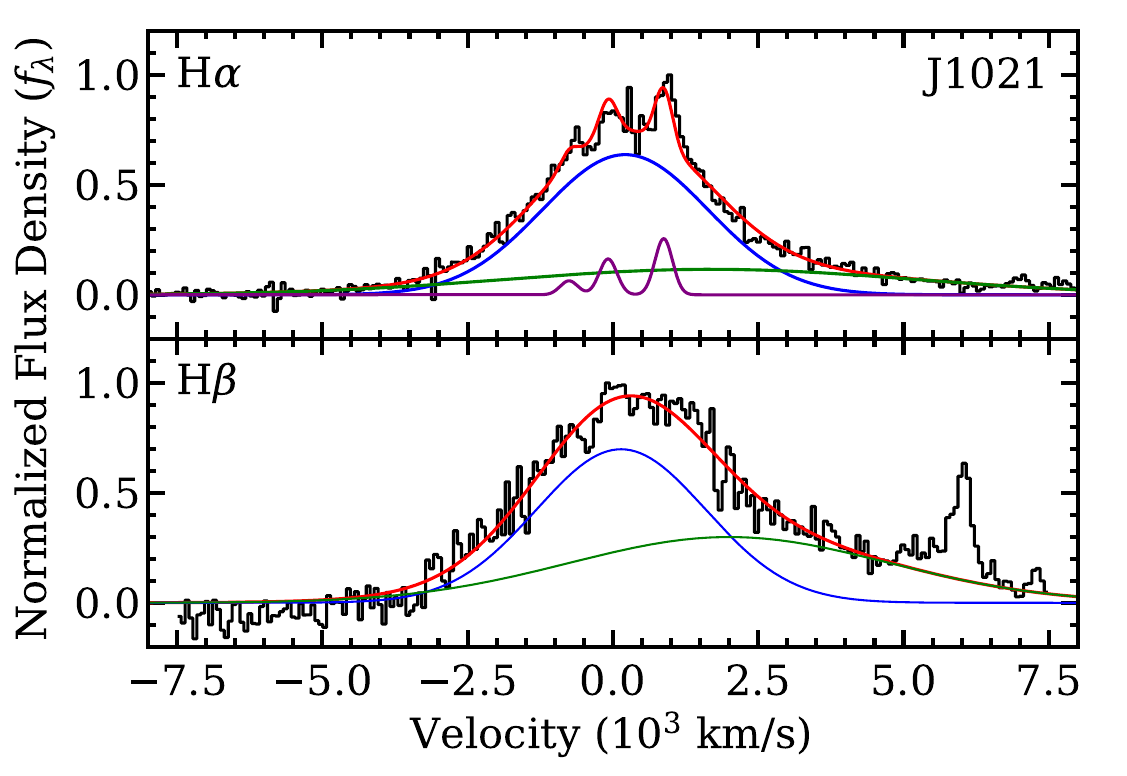}
\includegraphics[width=0.68\columnwidth]{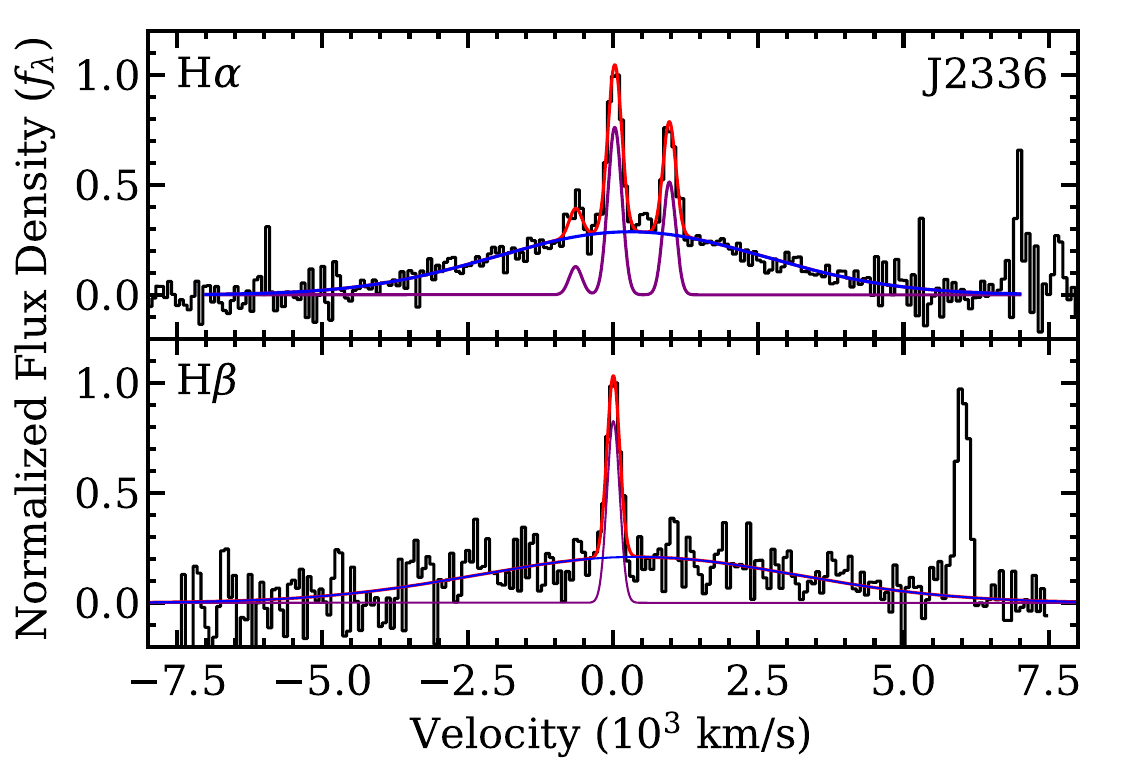}

\caption{High-state H$\alpha$ and H$\beta$ emission line profiles for J1011, J1021, and J2336 respectively. We show the continuum subtracted spectrum in black, the total best-fit emission line profile in red, and various components of the broad fit in blue in green. For high-state narrow H$\alpha$ and [N II], we use low-state fits to constrain the shapes, location, and amplitude of the high-state narrow lines. The narrow line complex is shown in purple.}
\label{fig:hi_fits}

\end{figure*}

We show the best-fit spectral decomposition for each quasar in Figure \ref{fig:spec_fits}, as well as the best emission line fits in the region around Ly$\alpha$, H$\alpha$ and H$\beta$ in the low state in Figure \ref{fig:emission_fits} and in the high state in Figure \ref{fig:hi_fits}. We report measured fluxes in the broad emission lines in Table \ref{table:meas}, and measured fluxes in the low-state narrow lines in Table \ref{table:low}.

\begin{deluxetable*}{ccccccccccccc}
\rotate
\tablecaption{CLQ Measured Properties \& Fluxes}
\tablewidth{0pt}
\label{table:meas}
\setlength{\tabcolsep}{4pt}
\tablehead{
{} & {} & {5100 \AA{}} & {} & {} & {Broad} & {Broad} & {Broad} & {Broad} & {Broad} & {Broad} & {Total} & {Total}\\
{Object} & {State} & {Continuum Flux} & {$\alpha_{ox}$} &{$\eta_\mathrm{Edd}$} & {H$\alpha$ Flux} & {H$\alpha$ FWHM} &{H$\beta$ Flux} & {H$\beta$ FWHM} & {Ly$\alpha$ Flux} & {Ly$\alpha$ FWHM} &{Ly$\alpha$ Flux} & {Ly$\alpha$ FWHM}\\
{(1)} & (2) & (3) & (4) & (5) & (6) &(7) & (8) & (9) & (10) & (11) & (12) & (13)}
\startdata
{J1011} & {\textit{bright}} & {10.7 $\pm$ 0.2} & {\dots} & {0.13} & {4740 $\pm$ 40} & {2210 $\pm$ 30 \,} & {1060 $\pm$ 30} & {2090 $\pm$ 210} & {\dots} & {\dots} & {\dots} & {\dots}\\
{} & {\textit{faint}} & {0.80 $\pm$ 0.03} & {1.7} & {0.01} & {\, 180 $\pm$ 10} & {2900 $\pm$ 140} & {\, 25 $\pm$ 4} & {2900 $\pm$ 140} & {\, 80 $\pm$ 20} & {1330 $\pm$ 390} & {190 $\pm$ 20} & {1070 $\pm$ 170}\\
{} & {} & {} & {} & {} & {} & {} & {}\\
{J1021} & {\textit{bright}} & {11.3 $\pm$ 0.2} & {\dots} & {0.020} & {4700 $\pm$ 30} & {3340 $\pm$ 50 \,} & {1250 $\pm$ 20} & {7280 $\pm$ 570} & {\dots} & {\dots} & {\dots} & {\dots}\\
{} & {\textit{faint}} & {3.0 $\pm$ 0.3} & {1.1} & {0.009} & {\, 460 $\pm$ 20} & {5720 $\pm$ 180} & {\, \, 60 $\pm$ 10} & {5720 $\pm$ 180} & {1490 $\pm$ 190} & {\, 5610 $\pm$ 1300} & {2390 $\pm$ 130} & {1420 $\pm$ 170}\\
{} & {} & {} & {} & {} & {} & {} & {}\\
{J2336} & {\textit{bright}} & {1.1 $\pm$ 0.1} & {\dots} & {0.005} & {320 $\pm$ 10} & {5560 $\pm$ 210} & {\, 70 $\pm$ 10} & {6320 $\pm$ 480} & {\dots}  & {\dots} & {\dots}  & {\dots}\\
{} & {\textit{faint}} & {0.80 $\pm$ 0.01} & {1.6} & {0.003} & {110 $\pm$ 10} & {6540 $\pm$ 410} & {$<20$} & {6540 $\pm$ 410} & {\, 60 $\pm$ 20} & {1830 $\pm$ 850} & {140 $\pm$ 20} & {870 $\pm$ 30}\\
\enddata
\tablecomments{Column 1: SDSS object name, Column 2: The state of the quasar at the time of the measurements. Bright refers to states with strong continuum and broad emission, faint refers to states with weak broad emission, Column 3: Measured 5100 \AA{} quasar continuum flux from decomposed spectrum in units $10^{-17}$ erg s$^{-1}$ cm$^{-2}$ \AA{}$^{-1}$, Column 4: $\alpha_{ox}$ is calculated using the equation from \cite{tananbaum79}. We use an extrapolated $2500$ \AA{} luminosity and the X-ray fluxes reported in \cite{ruan19} to construct the value, Column 5: Calculated Eddington ratio, ($\frac{L_{bol}}{L_{Edd}}$). Literature black hole masses are used to find $L_{Edd}$, and $L_{bol}$ is calculated following the procedures described in Section \ref{sec:edd} of the text. In short, high-state values are calibrated using the \cite{runnoe12} 5100~\AA\ bolometric correction, and low-state values are constructed using the shape-dependent \cite{lusso10} bolometric correction, Columns 6, 8, 10, 12: Broad emission line fluxes in units $10^{-17}$ erg s$^{-1}$ cm$^{-2}$, Columns 7, 9, 11, 13: Emission line FWHM in units of km~s$^{-1}$. Columns 10, 11: Emission line flux and FWHM from the broad base of Ly$\alpha$ in units of $10^{-17}$ erg s$^{-1}$ cm$^{-2}$ and km~s$^{-1}$, Columns 12, 13: Total Ly$\alpha$ flux and FWHM (base + core of emission line) in units of $10^{-17}$ erg s$^{-1}$ cm$^{-2}$ and km~s$^{-1}$.} 
\end{deluxetable*}

\begin{deluxetable*}{cccccccccc}
\tablecaption{Low-State Narrow Line Fluxes}
\tablewidth{0pt}
\label{table:low}
\setlength{\tabcolsep}{5pt}
\tablehead{
{} & {[O II]} & {H$\beta$} & {[O III]} & {[O I]} & {[O I]} & {H$\alpha$} & {[N II]} & {[S II]} & {[S II]}\\
{Object} & {$\lambda 3727$} & {$\lambda 4861$}& {$\lambda\lambda 4959, 5007$} & {$\lambda 6300$} & {$\lambda{6363}$} & {$\lambda6563$} & {$\lambda\lambda 6548, 6583$} & {$\lambda 6716$} & {$\lambda 6731$}\\
{(1)} & (2) & (3) & (4) & (5) & (6) &(7) & (8) & (9) & (10)}
\startdata
{J1011} & {58 $\pm$ 14} & {4 $\pm$ 2} & {115 $\pm$ 6} & {\dots} & {6 $\pm$ 3} & {29 $\pm$ 5} & {49 $\pm$ 8} & {21 $\pm$ 3} & {13 $\pm$ 6}\\
{} & {} & {} & {} & {} & {} & {} & {}\\
{J1021} & {60 $\pm$ 13} & {26 $\pm$ 13} & {400 $\pm$ 45} & {26 $\pm$ 13} & {64 $\pm$ 7} & {145 $\pm$ 8} & {278 $\pm$ 13} & {66 $\pm$ 9} & {56 $\pm$ 7}\\
{} & {} & {} & {} & {} & {} & {} & {}\\
{J2336} & {35 $\pm$ 6} & {11 $\pm$ 1} & {78 $\pm$ 4} & {5 $\pm$ 2} & {\dots} & {45 $\pm$ 2} & {36 $\pm$ 2} & {10 $\pm$ 3}  & {11 $\pm$ 1}\\
\enddata
\tablecomments{Column 1: SDSS object name, Columns 2-10: Narrow line fluxes in units of $10^{-17}$ erg s$^{-1}$ cm$^{-2}$. Narrow line flux measurements were made on low-state HET spectra, except in the case of the [S II] doublets in J1011 and J2336. In the HET spectra, the telluric band corrections were not precise enough to measure [S II] for J1011 or J2336, so those fluxes were measured from the low-state SDSS spectra.} 
\end{deluxetable*}

\subsection{UV Spectral Decomposition}\label{sec:uv}

We only had low-state UV observations of the three CLQs from the HST. In all three spectra, there were discernible emission lines in the G140L grating, and the continuum was weak and smooth. We first fit the UV continuum with a power law, then fit the emission lines with a combination of Gaussians. For two of the quasars, the only emission line that we detected was Ly$\alpha$. Thus, in both J1011 and J2336, Ly$\alpha$ was well described with a combination of two Gaussians. In the case of J1021, we resolved both more structure in the Ly$\alpha$ emission line and a much larger set of emission lines, including N V, Si IV, and C IV.  We therefore fit Ly$\alpha$ with a combination of three Gaussians, N V with two Gaussians, Si~IV with a single Gaussian, and C IV with two Gaussians. We show the best emission line fits for Ly$\alpha$ in each quasar in the top panels of Figure \ref{fig:emission_fits}. We report the flux and other properties of Ly$\alpha$ in Table \ref{table:meas}.

The SNR in all G230L spectra was low and no emission lines were discernible (excepting J1021, where only C~IV was observed). Therefore, we combined the COS G140L and G230L observations in order to describe the shape of the quasar continuum in the UV. To achieve this, we heavily binned the G140L and G230L observations and combined them with the optical spectra in order to construct the spectral energy distribution from the near-UV to the near-IR. We used four bins in the G140L (avoiding emission lines) and one bin in the G230L bandpasses for this step.

\section{Spectral Fitting and Measurements for Individual Objects}\label{sec:indiv}

\subsection{J1011}
\begin{figure}
\centering
\includegraphics[width=\columnwidth]{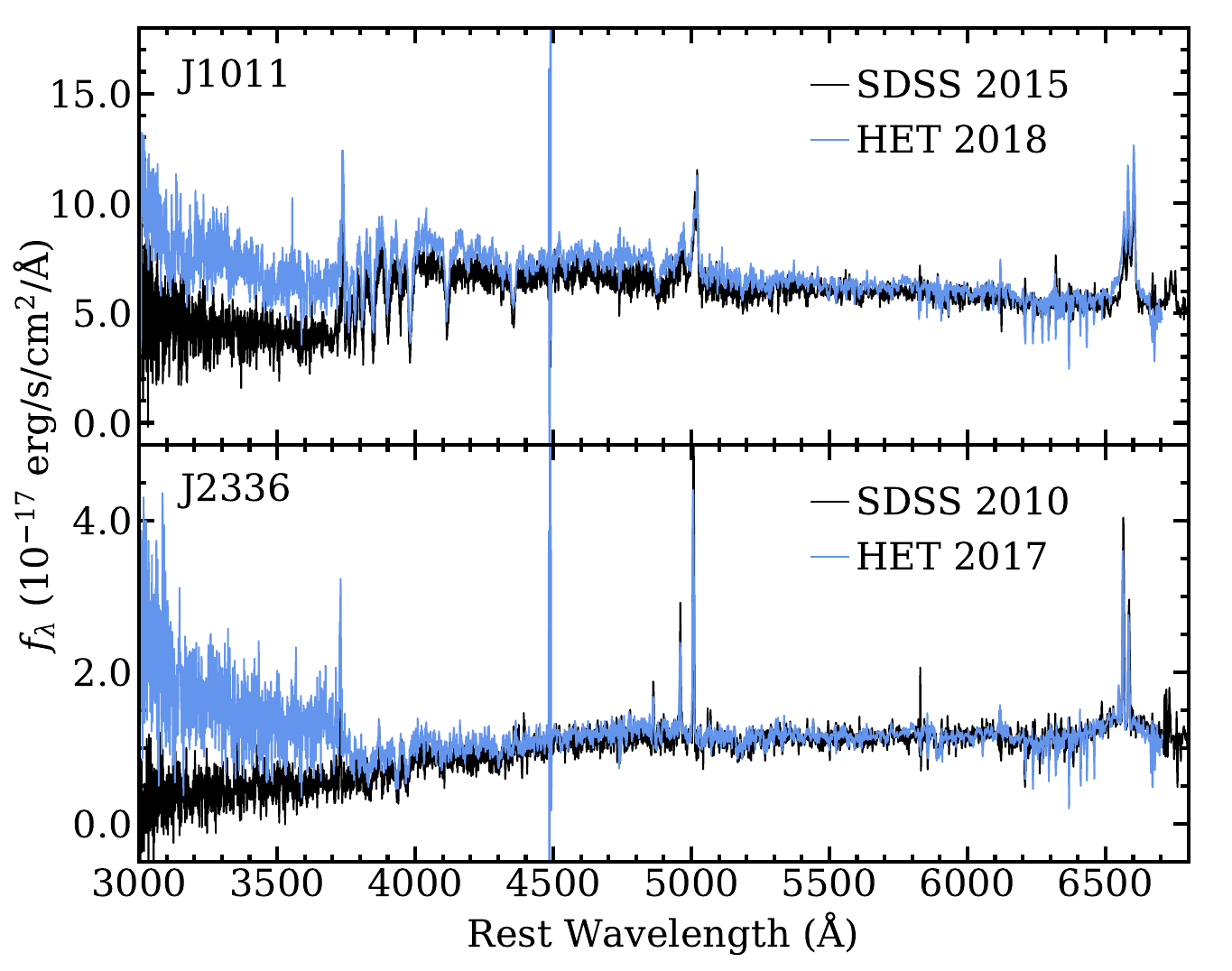}
\caption{\textit{Top}: J1011's spectral evolution since 2015. In the light blue is the HET spectrum from 2018 and in the black is the SDSS low-state spectrum from 2015. Since 2015, the blue continuum in J1011 has begun to grow stronger. \textit{Bottom}: J2336's spectral evolution since 2010. In the light blue, again, is the HET spectrum from 2017 and in the black is the SDSS low-state spectrum from 2010. In the seven years between observations, the blue continuum in J2336 has also begun to increase in strength. J1021 is the only CLQ in this sample that does not show this behavior.}
\label{fig:spec_ev}
\end{figure}

In J1011, the low state had broad H$\alpha$ emission that was well described by one narrow component, two narrow [N II] Gaussians and one broad Gaussian component, that was offset from narrow H$\alpha$ by $\sim 100$~km~s$^{-1}$. We used the narrow H$\alpha$ location and FWHM to constrain the narrow H$\beta$. We found that there was potentially some small contribution from broad H$\beta$, and fit H$\beta$ with a combination of a broad and narrow Gaussian. Of note is that the low state of J1011 has shown some evolution since the time that the SDSS low-state observation was taken -- the blue continuum appears to have risen between the first low-state spectrum taken in 2015 \citep{runnoe16} and the HET spectrum taken in 2018, although we did not find a significant change in the low-state broad H$\alpha$ and H$\beta$ emission line fluxes between the observation reported in \cite{runnoe16} and now (see Figure \ref{fig:spec_ev}, top for evolution in J1011's low state).

Ly$\alpha$ here was well fit by a combination of two Gaussians. Combined, the two Gaussians had a FWHM of 1065~km~s$^{-1}$ and the broad base Gaussian had a FWHM of 1331~km~s$^{-1}$. Both the base and the base plus core had significantly broader FWHMs than the narrow Balmer lines, which had a FWHM of $\sim 330$~km~s$^{-1}$. We also found that the UV continuum rose towards longer wavelengths, whereas the optical continuum grew stronger towards shorter wavelengths.

We used the strength of the low-state narrow lines to constrain the narrow lines in the high-state. We then fit two broad components to both H$\alpha$ and H$\beta$ in combination with the low-state narrow line constraints. The broad component of the high-state Balmer lines, shown in blue in Figure \ref{fig:hi_fits}, had very similar central wavelengths as the broad Balmer components in the low state.

\subsection{J1021}
\begin{figure}
\centering
\includegraphics[width=\columnwidth]{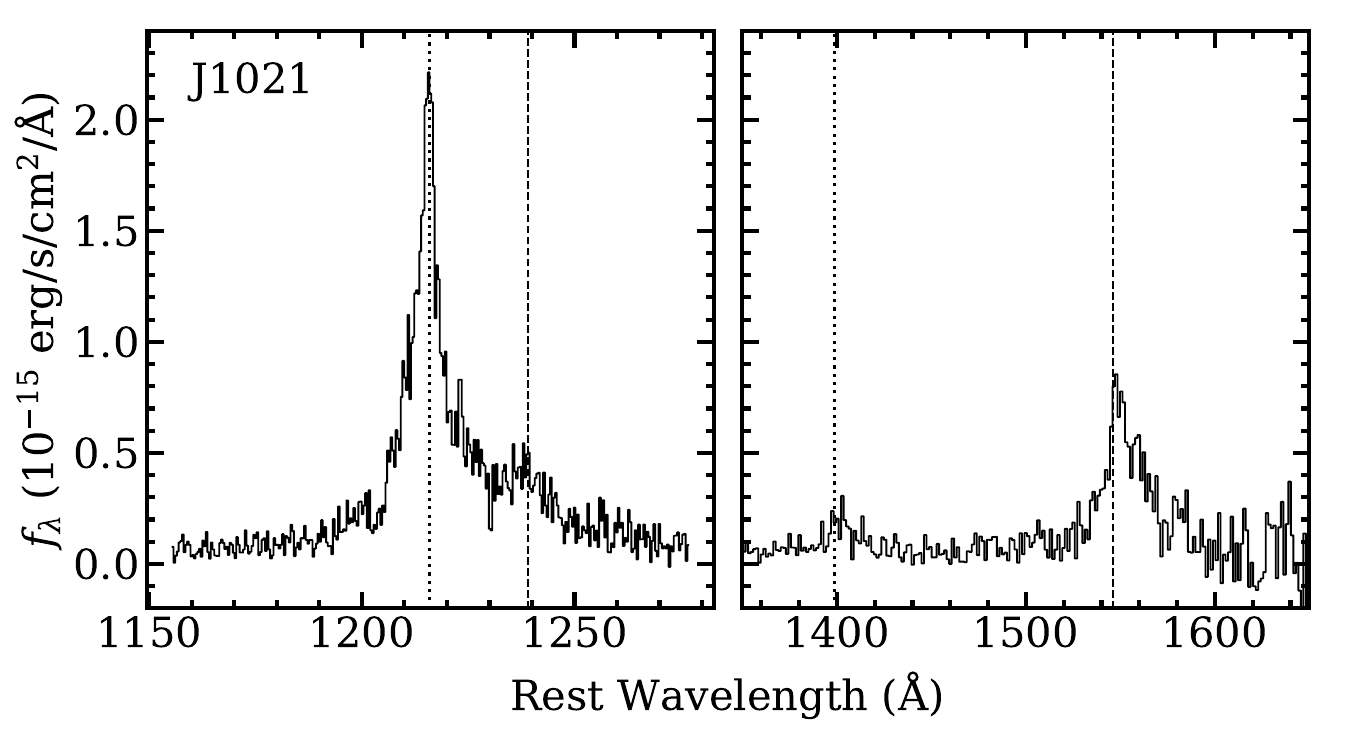}
\caption{\textit{Left}: Ly$\alpha$ and N~V emission lines in J1021's low state. Nominal wavelengths of Ly$\alpha$ and N~V are marked by the dotted and dashed lines respectively. We note that N~V is stronger than typical quasar composite spectra would suggest \citep{vandenberk01}. \textit{Right}: Si~IV and O IV] blended and C~IV emission in J1021's spectrum. The nominal wavelength of the Si~IV O IV] blend is marked with the dotted line, and the nominal wavelength of C~IV is marked with the dashed line.}
\label{fig:j1021_lya}
\end{figure}

The low-state H$\alpha$ emission line of J1021 was well described by a combination of one narrow H$\alpha$ component, two narrow [N II] components, and a broad H$\alpha$ Gaussian component that is redshifted from the narrow H$\alpha$ by $500$~km~s$^{-1}$. Unfortunately, the spectral region around H$\beta$ was contaminated by telluric Na~I~D emission. We were still able to fit narrow H$\beta$ and a weak broad H$\beta$ component, but we masked out the Na~I~D residual, which reduced the confidence of our fit because the contamination spanned a range $\sim 500$--2000~km~s$^{-1}$ from the center of the narrow line. This was the only CLQ for which N~V, Si~IV, and C~IV are also discernible, as we illustrate in Figure \ref{fig:j1021_lya}. J1021 did not show significant evolution between the SDSS low-state spectrum, taken in 2014, and the HET low-state spectrum, taken in 2018.


In the high state, J1021's broad Balmer emission was well fit with two Gaussians. We constrained the strengths of the narrow H$\alpha$ and [N~II] lines in the high state to be the same as they were in the low state. The broad components for H$\beta$ and H$\alpha$ in the high state had very similar central wavelengths. 

\subsection{J2336}
We used one narrow H$\alpha$ component, two narrow [N~II] Gaussians, and a broad H$\alpha$ Gaussian component to characterize the low-state H$\alpha$ emission in J2336. The narrow H$\alpha$ was blueshifted from broad H$\alpha$ by $\sim550$~km~s$^{-1}$. In J2336, the broad H$\beta$ was undetectable. The fit we show in Figure \ref{fig:emission_fits} and the flux we report for broad H$\beta$ in Table \ref{table:meas} represent upper limits for one broad component.

Ly$\alpha$ emission in the low state in J2336 was noticeably broader than the narrow lines but much narrower than the broad H$\alpha$ component that we detected in the high state. In this quasar, the Ly$\alpha$ emission line had a FWHM of $\sim900$~km~s$^{-1}$, whereas the narrow Balmer lines have a FWHM $\sim300$~km~s$^{-1}$. We used two Gaussians to fit the observed profile, which is not symmetric but rather has a red tail.

In the high state, J2336 showed relatively weak broad H$\alpha$ and broad H$\beta$. Both lines were best fit with single broad components that had similar FWHM (5560~km~s$^{-1}$ and 6320~km~s$^{-1}$ respectively). 

J2336 also showed some evolution in the low state since initial identification, $\sim2500$~days before follow-up HET spectra were taken (see Figure \ref{fig:spec_ev}, bottom). As in J1011, the blue continuum appears to have become steeper, although there has not been an appreciable change in the broad H$\alpha$ or H$\beta$ emission strength.

\section{Analysis}\label{sec:analysis}
\subsection{Dust Extinction}\label{sec:dustext}
In order to fully test the variable obscuration hypothesis, we constructed a range of emission line ratios and used the expected values of these emission line ratios to infer V-band extinction (A$_\mathrm{V}$) values. If variable dust obscuration is indeed the source of the change, the calculated A$_\mathrm{V}$ values for a single quasar should be self-consistent, regardless of which method is used to construct the value. To calculate A$_\mathrm{V}$, we first assumed a \cite{ccm89} extinction law with R$_\mathrm{V}~=~3.1$. 

For both the high and low states, we measured A$_\mathrm{V}$ from H$\alpha$/H$\beta$ ratios. To do this, we ignored any narrow emission lines and assumed that the `intrinsic' H$\alpha$/H$\beta$ value for the broad line region was that measured from the high state, and that any change in the ratio we observed in the low state was entirely due to dust extinction. We also measured A$_\mathrm{V}$ from H$\alpha$ and H$\beta$ broad flux measurements alone, again assuming that the true, unextinguished flux for each of the broad lines was the value we observed in the high state, and that the drop in flux we saw in the low state was due to dust. We further utilized the Ly$\alpha$/H$\alpha$ ratio, although we only had low-state measurements of Ly$\alpha$. As such, we adopted a nominal high-state Ly$\alpha$/H$\alpha \approx 2-5$ \citep[see][]{vandenberk01, k02, tang12}. Finally, we measured the power-law continuum strength at 5100~\AA\ in both the high and low states, and attributed any dimming in the continuum strength in the low state entirely to dust. We report derived A$_\mathrm{V}$ values for each method for each CLQ in Table \ref{table:av}. 

Various dust attenuation and extinction laws are very similar in the optical, but can diverge wildly in the UV. To account for the range in UV attenuation associated with different attenuation laws, we calculated `maximum' and `minimum' values of A$_\mathrm{V}$ from the Ly$\alpha$/H$\alpha$ ratio by using both the \cite{calzetti94} attenuation law for starburst galaxies, which is greyer than most attenuation laws, and the Small Magellanic Cloud attenuation law \citep{gordon03}, which is steeper into the UV than most attenuation laws. This range, created by varying which attenuation law we use to calculate A$_\mathrm{V}$, is reflected in the error bars reported in Table \ref{table:av}.

\begin{figure}
    \centering
    \includegraphics[width=\columnwidth]{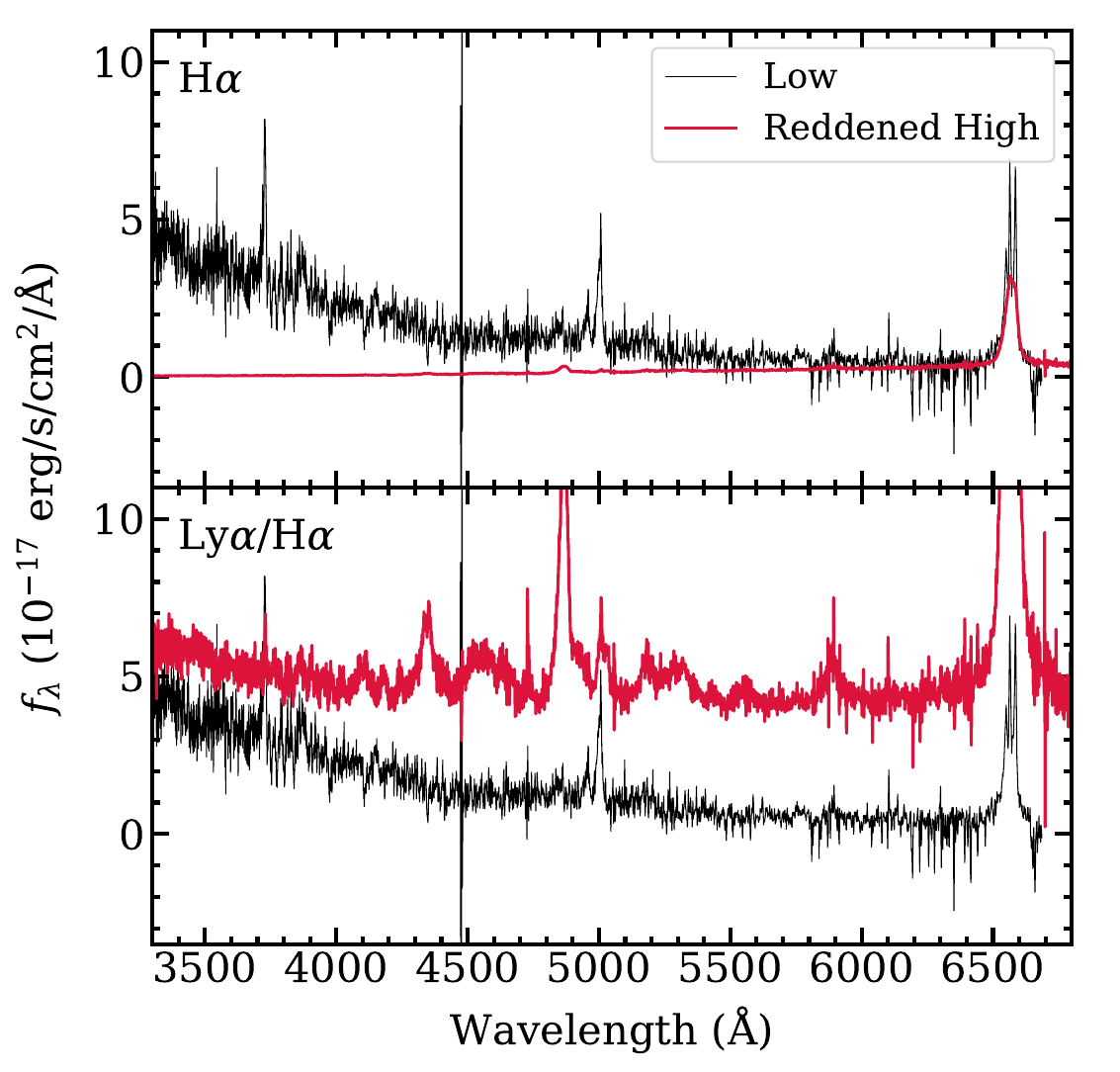}
    \caption{In both panels, we show in red the high-state spectrum of J1011 with all stellar contribution removed and reddened to the level of the low-state spectrum (shown in black, also with all stellar contribution removed). The high-state spectrum was taken by the SDSS and the low-state spectruum was taken with the HET. In the top panel, we have reddened the high-state spectrum using the \cite{ccm89} extinction law with an R$_\mathrm{V}$ of 3.1 and with $\mathrm{A}_\mathrm{V} = 4.39$, calculated from the broad H$\alpha$ fluxes in the high and low state. We see that, despite the fact that broad H$\alpha$ appears to match quite well between the reddened high state and the low state, the rest of the continuum and the broad H$\beta$ line do not match. In the lower panel, we perform the same test, but with $\mathrm{A}_\mathrm{V} = 1.00$, the value calculated from the Ly$\alpha$/H$\alpha$ ratio. In this case, reddening alone cannot account for any aspect of the observed change between the high and low state.}
    \label{fig:atten}
\end{figure}

Crucially, for all three quasars, the A$_\mathrm{V}$ values calculated in the optical and those calculated in the UV differed dramatically from each other. For J1011, none of the calculated A$_\mathrm{V}$ values were self-consistent. For J1021 and J2336, A$_\mathrm{V}$ values measured from optical emission lines agreed within error bars, but the optical continuum level and the Ly$\alpha$/H$\alpha$ values did not agree with the values calculated using optical emission lines. Thus, dust obscuration alone cannot explain the change we see, although a combination of variable dust extinction with some other mechanism that dims the continuum, such as a variable accretion rate, cannot be ruled out. To illustrate that dust attenuation alone is not a viable solution, in Figure \ref{fig:atten} we show the low-state spectrum of J1011 compared to its reddened high-state spectrum. J1011's low-state spectrum could not be replicated by simply reddening the high-state spectrum to the level of the broad H$\alpha$ emission or the Ly$\alpha$/H$\alpha$ ratio in the low state. 

\begin{deluxetable*}{cccccccc}
\tablecaption{Calculated A$_\mathrm{V}$ Values}
\tablewidth{0pt}
\label{table:av}
\setlength{\tabcolsep}{10pt}
\tablehead{
{} &{} & {} & {} & {Ly$\alpha$/H$\alpha$} & {Ly$\alpha$/H$\alpha$} & {5100 \AA{}}\\
{Object} &{H$\alpha$} & {H$\beta$} & {H$\alpha$/H$\beta$} & {base} & {base + core} & {Continuum}\\
{(1)} & (2) & (3) & (4) & (5) & (6) & (7)}
\startdata
{J1011} & {4.4 $\pm$ 0.1} & {3.6 $\pm$ 0.2} & {1.6 $\pm$ 0.6} & {1.0 $\pm$ 0.3} & {0.6 $\pm$ 0.1} & {2.6 $\pm$ 0.1}\\
{J1021} & {3.1 $\pm$ 0.1} & {2.9 $\pm$ 0.2} & {2.3 $\pm$ 0.8} & {0.2 $\pm$ 0.1} & {0.01 $\pm$ 0.01} & {1.3 $\pm$ 0.1}\\
{J2336} & {1.4 $\pm$ 0.1} & {$>$1.2} & {$>$0.9} & {0.9 $\pm$ 0.1} & {0.6 $\pm$ 0.1} & {0.4 $\pm$ 0.1}\\
\enddata
\tablecomments{Column 1: SDSS object name, Column 2 -- 7: Inferred A$_\mathrm{V}$ values from high and low state fluxes, assuming a \cite{ccm89} extinction law. Error bars include 1$\sigma$ standard deviations derived from bootstrap resampling and include variation between different extinction laws (see Section \ref{sec:dustext} for details).} 
\end{deluxetable*}

\subsection{Low-State UV-Optical Spectral Energy Distribution}\label{sec:sed}
\begin{figure}
    \centering
    \includegraphics[width=\columnwidth]{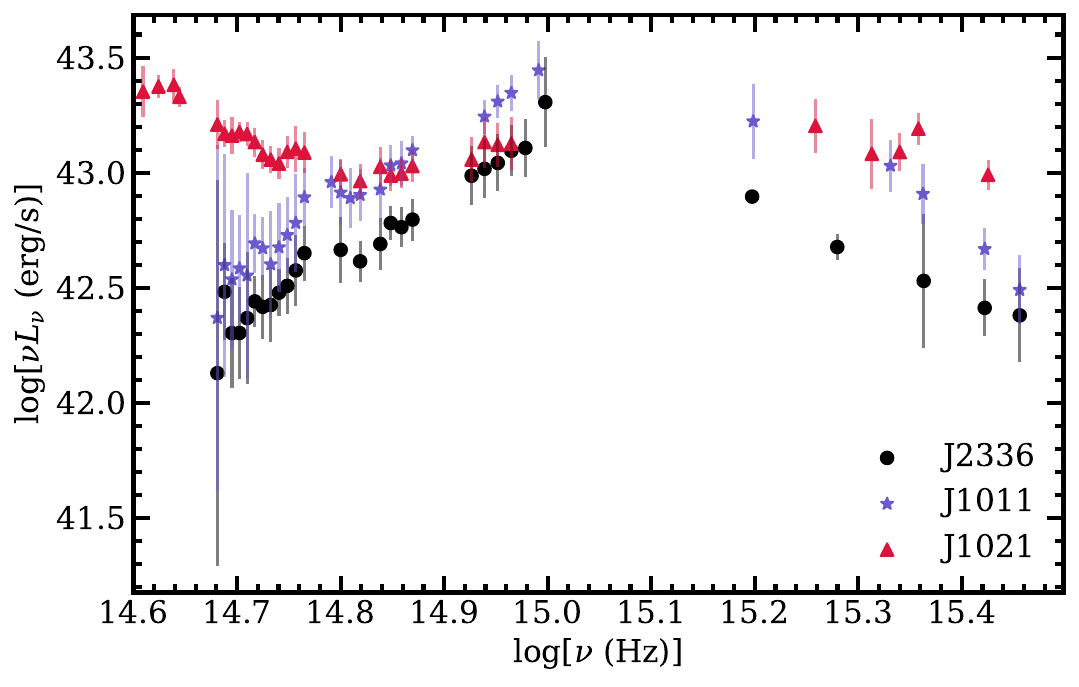}
    \caption{The (non-stellar) spectral energy distribution for each low-state CLQ. We show J2336 in the black circles, J1011 in the blue stars, and J1021 in the red triangles. J1011 and J2336 have nearly identical SED shapes, but J1021 differs at low frequency from the other two. Error bars here represent the standard deviation of the flux density of individual pixels in each bin (see Section \ref{sec:sed} for more detail). }
    \label{fig:clq_seds}
\end{figure}

We used the low-state HST and HET spectra to construct UV-optical SEDs for each quasar in the low-state. In each low-state HET optical spectrum, we coarsely binned the continuum after subtracting the starlight contribution, avoiding the regions around emission lines. In each low-state HST UV spectrum, we coarsely binned the G140L and G230L spectra, again avoiding the regions around Ly$\alpha$ and C IV. For all three quasars, we show the resulting $\nu L_\nu$ values plotted against the frequency in Figure \ref{fig:clq_seds}. 

\begin{figure}
    \centering
    \includegraphics[width=\columnwidth]{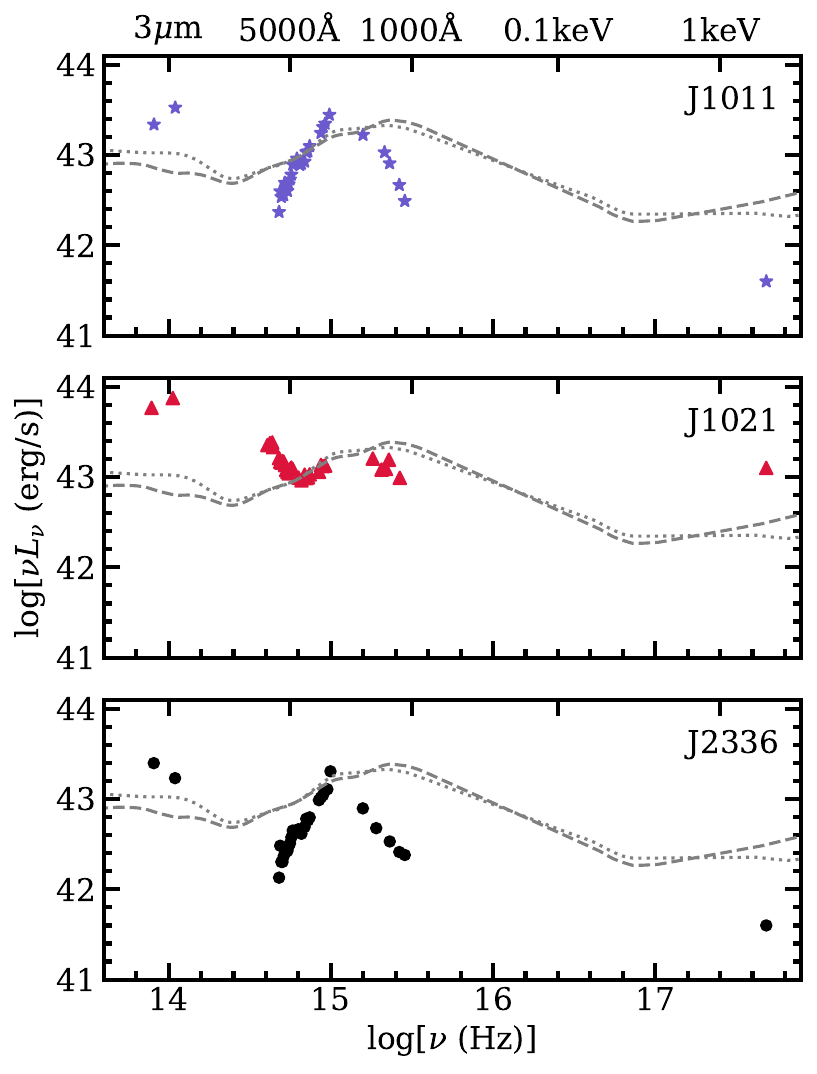}
    \caption{Full IR through X-ray low-state SEDs of J1011 (top), J1021 (middle), and J2336 (bottom). Overplotted in the grey dashed and dotted lines are \cite{elvis94} composite radio loud and radio quiet quasar SEDs. The IR data are drawn from NEOWISE and X-ray data come from Chandra \citep{ruan19}. None of the SEDs peak in the same place as composite quasar templates.}
    \label{fig:full_seds}
\end{figure}

We also show IR through X-ray SEDs of our CLQs in Figure \ref{fig:full_seds}, constructed with our data in the optical and UV, and supplemented with the nearly contemporaneous X-ray fluxes reported in \cite{ruan19} and with contemporaneous IR data from NEOWISE in the W1 and W2 filters \citep{neowise}. \textit{Chandra measurements} reported by \cite{ruan19} were obtained while the CLQs were confirmed to be in their low states, and were taken between 51--177 days from the HET spectra. The NEOWISE measurements were taken while the CLQs were in their low states between $3-43$ days from the date of the HET low-state spectra. The WISE W1 and W2 bands have PSF FWHM values that are $\sim 6\farcs5$ in both bands, and the pixel scale is $\sim 2\farcs8$~pixel$^{-1}$. This means that the  NEOWISE photometry includes the entire host galaxy, not just the quasar. Thus, the IR fluxes shown in the SED can be considered an upper limit on the IR flux from the quasar component of the CLQ. We overplot the \cite{elvis94} radio loud and radio quiet composite quasar SEDs in the grey dashed and dotted lines respectively. Neither the IR-X-ray SEDs nor the optical-UV SEDs resemble traditional composite quasar SEDs.

\begin{figure}
    \centering
    \includegraphics[width=\columnwidth]{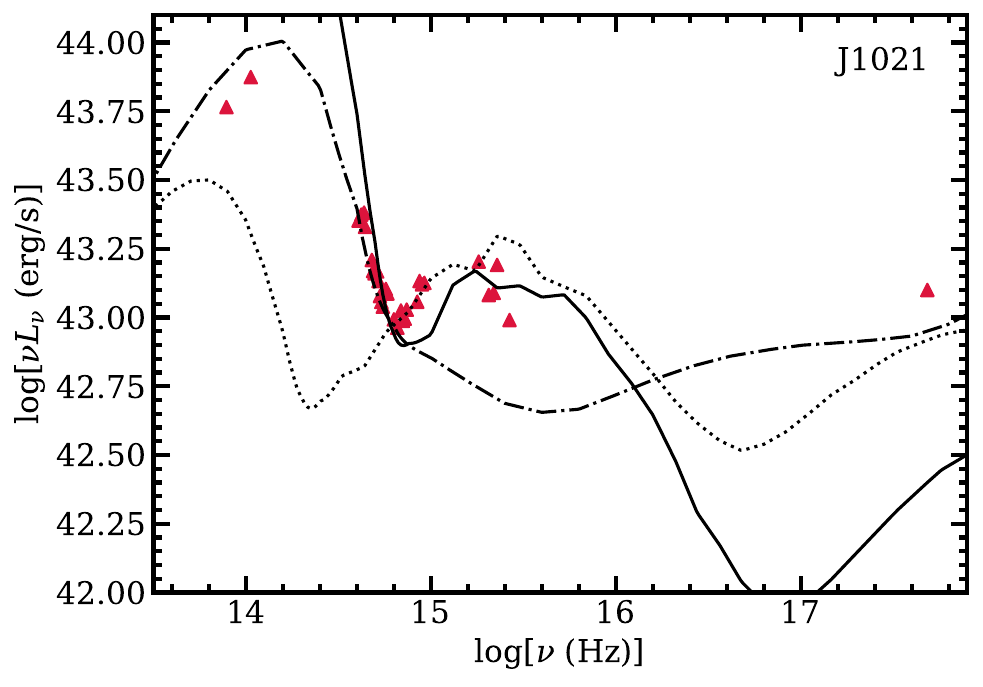}
    \caption{Full IR through X-ray SED of J1021. Overplotted in the black solid, dot-dashed and dotted lines are three different ADAF+truncated thin disk fits to LLAGN from \cite{nemmen14} and scaled to the luminosity at log($\nu/\mathrm{Hz})=14.8$. The \cite{nemmen14} fits were performed for LLAGN with significantly lower Eddington ratios than the low-state CLQs we investigate here, but their SEDs are qualitatively more similar to the low-state CLQs than composite quasar SEDs.}
    \label{fig:1021_seds}
\end{figure}

For each of the three quasar SEDs, the continuum has vastly different slopes in the UV and the optical. In all three cases in the low state, the UV slope rises towards longer wavelengths while the optical quasar continuum rises towards short wavelengths. All three quasars lack the canonical UV bump that is commonly present in quasars. The UV bump is frequently associated with emission from a geometrically thin, optically thick accretion disk \citep[see][]{frank02}, and is visible in typical composite quasar spectra, regardless of where the sample is drawn from \citep[e.g.][]{elvis94, scott04, richards06, shang11}. Our sample has notably different slopes and shapes, and completely lacks the peak in the UV seen in those quasar templates. Even individual quasars in those samples do not show the shapes we see. We show a comparison between the SED of J1021 and three different ADAF+truncated thin disk model fits from \cite{nemmen14} for three different LLAGN in Figure \ref{fig:1021_seds}. None of the model fits shown here exactly match J1021, which has an Eddington ratio a few orders of magnitude higher than the LLAGNs fit in that work, but the shapes of those models and the flexibility in where they peak depending on physical properties of the AGN are qualitatively similar to the low-state SEDs we see here. Fitting the CLQ SEDs with ADAF+truncated thin disk models may provide better results than typical quasar SED models.

Interestingly, J1011 and J2336 have nearly identical SEDs in this state, whereas J1021 differs in its low frequency behavior. J1011 and J2336 also have nearly identical X-ray luminosities at 2 keV, whereas J1021 is almost two orders of magnitude brighter. The difference in SED shape between J1021 and the other two quasars thus extends from the optical all the way through to the X-ray.

Although J1011 and J2336 are nearly identical in this state, the short wavelength continuum is \textit{steeper} in the low state for J2336 than it was in the high state. This is somewhat unusual, and the only place that this happened in this sample.

\subsection{Eddington Ratio}\label{sec:edd}

We made use of 5100~\AA\ continuum measurements to calculate Eddington ratios in the high state. We used literature measurements of the black hole mass from \cite{runnoe16}, \cite{macleod16} and \cite{ruan16} in order to calculate $L_{Edd}$. In each of those works, black hole mass was calculated using the width of broad H$\beta$ and the continuum strength using the single epoch scaling relations from \cite{vp06}. For the calculation of $L_{bol}$, we first used the relationship defined in \cite{runnoe12} to convert a 5100~\AA\ luminosity to $L_{bol}$. This conversion yielded values in the high state that were consistent with values found in the literature for SDSS spectra in all three quasars. We report these high-state values in Table \ref{table:meas}. 

The SEDs of the three CLQs in the low-state do not resemble canonical quasar SEDs (see Section \ref{sec:disc} for discussion about the SED shape). This means that the conversion from a monochromatic luminosity to a total bolometric luminosity is less certain than if the SED resembled that of typical quasars. Thus, we calculated the low-state Eddington ratio using the SED shape-dependent bolometric corrections presented in \cite{lusso10}. This correction accounts for differences in the SED shape owing to X-ray contributions and depends on $\alpha_{ox}$, a ratio that defines the hardness of X-ray emission relative to a monochromatic luminosity at 2500~\AA\ \citep{tananbaum79}. We used the 2 keV X-ray luminosities, L$_{2\mathrm{keV}}$, reported in \cite{ruan19}, which were taken within 200 days of our spectra. We did not directly measure a luminosity at 2500~\AA, but instead extrapolated the UV trend seen in the HST spectra to 2500~\AA. We also extrapolated the optical quasar power-law to 2500~\AA. The values from extrapolation in both directions were consistent with each other, so we report $\alpha_{ox}$ calculated from the UV-extrapolated value in Table \ref{table:meas}. We then used the \cite{lusso10} bolometric corrections to construct Eddington ratios. The low-state values are reported in Table~\ref{table:meas}. 

We compared these calculated values to Eddington ratios constructed with other bolometric corrections, in the case that contemporaneous X-ray, UV, and optical measurements in the low-state are not available. When we calculated low-state Eddington ratios using the \cite{runnoe12} 5100~\AA\ correction that assumes a standard quasar SED, we found low-state Eddington ratios of 0.03, 0.01, and 0.003 for J1011, J1021, and J2336 respectively. These values are largely consistent with the shape-dependent Eddington ratios we found (0.01, 0.01, and 0.003 respectively). 

We also calculated Eddington ratios using bolometric corrections calibrated from LLAGN spectra in \cite{eracleous10}. There, the 2--10~keV X-ray luminosity, L$_{2-10\mathrm{keV}}$, is used to convert to a bolometric luminosity. For our measurement, we again used the 2 keV X-ray luminosities, L$_{2\mathrm{keV}}$, reported in \cite{ruan19}. We assumed a photon index $\Gamma = 1.8$ \citep[this value is commonly used for LLAGN; see][]{constantin09, gu09, younes11} to convert L$_{2\mathrm{keV}}$ to L$_{2-10\mathrm{keV}}$, and then used that luminosity to convert to a bolometric luminosity. The resulting Eddington ratios are, again, mostly in good agreement with the \cite{lusso10} corrections. J1011 and J2336 in particular had Eddington ratios with this method of 0.01 and 0.002, which are well in line with the other two methods. This method, however, yields an Eddington ratio for J1021 that is an order of magnitude higher than the other two methods -- this may be because this calibration does not account for differences in X-ray hardness, or because J1021's X-ray observation in \cite{ruan19} was taken $\sim 180$ days before our data were taken. X-ray emission in quasars varies on much shorter timescales than the BLR does, and the higher X-ray luminosity of J1021 may not be representative of the state the CLQ was in at the time of the optical and UV observations we present here.

 
In summary, we report the values calculated from the SED shape-dependent \cite{lusso10} bolometric corrections, but note that the values calculated using the 5100 \AA{} correction calibrated from traditional quasar spectra in \cite{runnoe12} agree quite well. The X-ray based LLAGN correction is mostly consistent with values from the other two methods, however for the one CLQ that is much brighter in the X-ray than the other two, the X-ray based method produces an implausibly high Eddington ratio.



\section{Discussion}\label{sec:disc}

We found that, in all three CLQs, dust attenuation was not a self-consistent explanation for the observed transitions. In J1011, optical emission line measurements alone were sufficient to rule out dust attenuation. For J1021 and J2336, 5100~\AA\ continuum and UV emission line measurements were also necessary to rule out the dust attenuation hypothesis. A$_\mathrm{V}$ values calculated from the optical emission lines were consistent within error bars in both J1021 and J2336, but the values derived from the 5100~\AA\ continuum and the values derived from the UV spectra differed significantly from the values derived from the optical emission lines. UV emission in particular provided much lower values for A$_\mathrm{V}$ than values derived from the optical spectrum. In all three cases, the presence of observable Ly$\alpha$ emission alone was enough to discredit the variable obscuration hypothesis.


We also found that the SEDs of all three CLQs in the low-state look very different from canonical quasar SEDs. In contrast to usual models and composite spectra, the low-state SEDs are reminiscent of both models and empirical measurements of the SEDs of LLAGN \citep[see][]{ho99, maoz07, eracleous10, nemmen14}. In models of LLAGN SEDs, the SED shape that we see here is reproduced by an ADAF component and a truncated thin disk component \citep{nemmen14}. The combination of the ADAF and the truncated thin disk allows the SED to peak in the IR and optical or NUV instead of the FUV. In those models, synchrotron and bremsstrahlung radiation are produced in the ADAF and then the synchrotron photons are upscattered via inverse Compton scattering. There are also components representing emission from the truncated thin disk. The peak in our low-state SEDs in all three cases occurs at log$(\nu/\mathrm{Hz})=15$--15.2. This peak can be explained by thermal emission from a truncated thin disk. Much like LLAGN, all three low-state CLQ SEDs peak at longer wavelengths than typical quasar SEDs and thus lack the UV bump that is present in typical quasar SEDs. The three optical-UV SEDs also appear to peak in around the same location. This suggests that the transition radius between the ADAF and the truncated thin disk may be comparable in all three cases.

Theoretical models of quasar accretion disks have found that the short timescale of CLQ state changes may be replicated via magnetic-pressure dominated disks with outflows \cite[e.g.][]{feng21,wugu23}. In our observations, however, we do not see evidence of fast moving outflows (i.e. blueshifted Ly$\alpha$ absorption in J1011 or J2336, or blueshifted C IV absorption in J1021) in either the high or low state. Additionally, the continuum SEDs that we measure do not peak in the same place that these models predict.

We also find that our UV spectra look qualitatively similar to other low-state spectra of both CLQs and extremely variable quasars (EVQs) in the UV. Past studies of UV-identified CLQs show spectra that have evolved over time and, in some low-state cases, show both the same lack of UV bump that we observe here and very weak C IV \citep[see][for examples]{guo20, guo24}. 

In all three CLQs, the Eddington ratio decreased between the high state and the low state. Our low-state SEDs differ from usual quasar SEDs, which means that the conversion from 5100~\AA\ luminosity to total bolometric luminosity using usual relationships for quasars may not be appropriate.  Thus, we used a shape-dependent conversion to yield a bolometric luminosity. When we used shape-dependent corrections \citep{lusso10} and bolometric corrections based on LLAGN SEDs in the X-ray \citep{eracleous10}, we found that, for the most part, the resulting Eddington ratios were consistent with the ratios calculated using typical quasar bolometric corrections. The sole exception was the X-ray based correction for J1021 in the low state, which produced an Eddington ratio five times \textit{higher} than the high state, despite the fact that the traditional quasar and UV to X-ray shape-dependent bolometric corrections were consistent with each other. This discrepancy in the X-ray calculated Eddington ratio for J1021 may be due to the short time-scale and large amplitude of X-ray variability seen in quasars or intrinsic differences in the hardness of X-ray emission from the AGN. The X-ray observation of J1021 as reported in \cite{ruan19} was taken $\sim180$ days before our data, and may not have been representative of the state of the CLQ when our UV and optical data were taken. Nonetheless, we find that, regardless of deviations in the low-state SEDs of CLQs from typical quasar SEDs, bolometric corrections based on composite quasar spectra and bolometric corrections based on LLAGN SEDs largely agree here. Quasar calibrated bolometric corrections may be sufficient for low-state CLQs, although more rigorous modeling of the SED would achieve more accurate measurements of the total bolometric luminosity.

Taken together, our data support the idea that the cause of the state transition in these three CLQs is likely a changing accretion rate causing a change in the accretion flow structure. Dust extinction alone cannot explain the change in the spectrum, and the low-state SED looks like LLAGN SEDs seen in the literature.

J1021 is somewhat unusual for the following reasons. In composite quasar spectra, N~V is weak compared to Ly$\alpha$ \citep[see][for composite quasar spectrum]{vandenberk01}, but the N V emission is stronger in J1021 than might generally be expected for a quasar (see Figure \ref{fig:j1021_lya}, left). Thus J1021 looks qualitatively similar to other CLQs in their low-state, where N~V appears to remain relatively strong despite declining C~IV and Ly$\alpha$ \citep[e.g.][]{ross20}. We also detect many more broad UV emission lines in J1021 than in J1011 or J2336 and it is also the quasar where we detect the most significant contribution from broad H$\beta$ in the low state. Literature values for the X-ray luminosity of J1021's low-state show it to be two orders of magnitude more luminous at 2 keV than the other two quasars in this sample \citep{ruan19}. Further, J1021 has the most massive black hole of all three quasars. The brighter X-ray luminosity may be caused by short-term fluctuations or flares in the X-ray. Because the X-ray luminosity is higher, the ADAF may illuminate the truncated thin disk further, causing the appearance of more emission lines in the UV.

Finally we notice that, in our sample, two of the three quasars have evolved since they first entered their low states.
Both J1011 and J2336 have shown some brightening in the time since their first entry to the low state (there were about 120 and 2500 days respectively between low-state SDSS and HET spectra). In contrast, J1021's low-state spectrum has remained remarkably steady in the 1500 days between the SDSS low-state and the HET low-state observation. This low-state evolution may be interesting to trace in a larger sample of turn-off CLQs. Indeed, some low states last for a long enough duration that even the narrow line emission changes \citep{yang24}.

\section{Summary and Conclusions}\label{sec:summary}
In this paper, we fit the UV-optical HST-HET spectra and constructed low-state SEDs of three CLQs that previously had prominent broad emission. We also investigated, in detail, one potential explanation for CLQs -- the variable dust obscuration hypothesis. Through our analysis, we found that:

\begin{enumerate}
    \item Dust extinction alone is insufficient to explain the change we saw in all three CLQs in our sample. In J1011, constraints from broad optical emission lines alone were enough to rule out the variable obscuration hypothesis. In J1021 and J2336, continuum measurements were required in conjunction with optical emission line measurements to rule out variable obscuration. In all three CLQs, the presence of Ly$\alpha$ in the low state corroborated and strengthened the above conclusions.
    \item All three CLQs displayed low-state SEDs that are remniscent of LLAGN SEDs. The SEDs peaked at lower frequency than typical quasar SEDs do, and appears to lack the UV bump that is a hallmark of typical quasar SEDs. This resemblance to LLAGNs strongly suggests that the cause of the observed state change may, in fact, be due to a large drop in accretion rate, leading to a transition from a thin accretion disk to an ADAF and a truncated thin disk.
    \item Two of the three CLQs show further evolution between the SDSS observed low state and the HET observation of the low state. While not a statistical sample, most of our quasars continue evolving and thus we posit that objects already identified as CLQs may show continued changes over time. It may be compelling to continue monitoring confirmed CLQs for extended periods of time \citep[see][for report of J1011's return]{yang24}. 
\end{enumerate}

To strengthen our conclusions, larger samples of CLQs must be studied in the same manner. Additional contemporaneous UV and optical spectra of CLQs in the low state, potentially in conjunction with X-ray observations, will allow for confirmation of our SED findings. Continued monitoring of already identified CLQs is also necessary to assemble a complete picture of the state change. UV and optical spectra in both the high and low states can be used to confirm whether the quasar SED truly changes its intrinsic shape between epochs. In addition, to confirm our findings, it would be beneficial to perform detailed SED modelling and fit LLAGN-type SEDs to the observed data. 

\bigskip
\begin{acknowledgments}
We thank the anonymous referee for their thoughtful comments that helped us improve the manuscript.
We thank Jules Halpern for carrying out the imaging observations at MDM Observatory and Yue Shen for his critical reading of the manuscript and his helpful comments.

Support for program GO-14799 was provided by NASA through a grant from the Space Telescope Science Institute, which is operated by the Association of Universities for Research in Astronomy, Inc., under NASA contract NAS5-26555. This work is supported by the Penn State Science Achievement Graduate Fellowship Program. LD acknowledges support from the Hasek Graduate Fellowship and the Graduate School Endowed Fellowship at Penn State, as well as the Pennsylvania Space Grant Consortium Graduate Fellowship.

The Low Resolution Spectrograph 2 (LRS2) was developed and funded by the University of Texas at Austin McDonald Observatory and Department of Astronomy, and by Pennsylvania State University. We thank the Leibniz-Institut fur Astrophysik Potsdam (AIP) and the Institut fur Astrophysik Goettingen (IAG) for their contributions to the construction of the integral field units. We acknowledge the Texas Advanced Computing Center (TACC) at The University of Texas at Austin for providing high performance computing, visualization, and storage resources that have contributed to the results reported within this paper.

This work is based on observations obtained at the MDM Observatory, operated by Dartmouth College, Columbia University, Ohio State University, Ohio University, and the University of Michigan.

This publication makes use of data products from the Near-Earth Object Wide-field Infrared Survey Explorer (NEOWISE), which is a joint project of the Jet Propulsion Laboratory/California Institute of Technology and the University of Arizona. NEOWISE is funded by the National Aeronautics and Space Administration.

This research made use of Photutils, an Astropy package for
detection and photometry of astronomical sources \citep{photutils}.

\end{acknowledgments}

%

\vspace{5mm}
\facilities{HST (COS), Sloan (BOSS), HET (LRS2), McGraw-Hill (Templeton), NEOWISE}


\software{astropy \citep{2013A&A...558A..33A,2018AJ....156..123A},  
          calcos \citep{coshand},
          costools \citep{coshand},
          extinction \citep{extinction},
          matplotlib \citep{matplotlib},
          pandas \citep{pandas},
          photutils \citep{photutils},
          pPXF \citep{pPXF},
          scipy \citep{scipy},
          specutils \citep{specutils}
          }




\bibliography{sample631}{}
\bibliographystyle{aasjournal}



\end{document}